\documentclass[9pt,journal,compsoc]{IEEEtran}
\usepackage{cite}
\ifCLASSINFOpdf
   \usepackage[pdftex]{graphicx}
   \graphicspath{{../pdf/}{../jpeg/}{../}}
  \DeclareGraphicsExtensions{.pdf,. jpeg,.png}
\else
   \usepackage[dvips]{graphicx}
  \graphicspath{{../eps/}}
   \DeclareGraphicsExtensions{.eps}
\fi
\usepackage[cmex10]{amsmath}

\usepackage{algorithmic}
\usepackage{array}
\usepackage{mdwmath}
\usepackage[tight,footnotesize]{subfigure}

\usepackage{pgf}
\usepackage{tikz}
\usepackage{amsfonts,amssymb}
\usetikzlibrary{arrows, decorations.pathmorphing, backgrounds, positioning, fit, petri, automata,shapes.geometric}
\usepackage{comment}

\begin{document}
%

\title{Modeling QoE of Video Streaming in Wireless Networks with Large-Scale Measurement of User Behavior}

\author{{Yuedong Xu, Zhujun Xiao, Hui Feng, Tao Yang, Bo Hu, Yipeng Zhou}

\thanks{A preliminary version was published in IEEE Globecom 2015 \cite{xiao_xu}.

Yuedong Xu, Zhujun Xiao, Hui Feng, Tao Yang and Bo Hu are with Department of Electronic Engineering, Fudan University, Shanghai, China. Yipeng Zhou is with School of Computer Science, Shenzhen University.

Email: \{ydxu, 13210720042, hfeng, taoyang, bohu\}@fudan.edu.cn, ypzhou@szu.edu.cn}}

\IEEEtitleabstractindextext{
\begin{abstract}
Unraveling quality of experience (QoE) of video streaming is very challenging in bandwidth shared wireless networks.
It is unclear how QoE metrics such as starvation probability and buffering time interact with dynamics of
streaming traffic load.
In this paper, we collect view records from one of the largest streaming providers in China over two weeks
and perform an in-depth measurement study on flow arrival and viewing time that shed light on the real traffic pattern.
Our most important observation is that the viewing time of streaming users fits a hyper-exponential distribution quite well.
This implies that all the views can be categorized into two classes, short and long views with separated
time scales. We then map the measured traffic pattern to bandwidth shared 
cellular networks and propose an analytical framework to compute the
closed-form starvation probability on the basis of ordinary differential equations (ODEs). Our framework can be
naturally extended to investigate practical issues including the progressive downloading and the finite video duration.
Extensive trace-driven simulations validate the accuracy of our models.
Our study reveals that the starvation metrics of the short and long views possess different sensitivities to the
scheduling priority at base station. Hence, a better QoE tradeoff between the short and long views has a potential to
be leveraged by offering them different scheduling weights. The flow differentiation involves tremendous technical and
non-technical challenges because video content is owned by content providers but not the network operators and the
viewing time of each session is unknown beforehand.
To overcome these difficulties, we propose an online Bayesian approach to infer the viewing time of each incoming flow
with the ``least'' information from content providers.
\end{abstract}

}

%

\maketitle
\IEEEdisplaynontitleabstractindextext
\IEEEpeerreviewmaketitle

\section{Introduction}

\IEEEPARstart{M}{obile} video service is witnessing a tremendous growth nowadays. As forecasted by Cisco,
mobile video will increase
13-folds between 2014 and 2019, accounting for 72\% of total
mobile data traffic by 2019 \cite{cisco}. In particular, video on demand
(VoD) dominates the streaming traffic on mobile Internet. The VoD has a less stringent requirement on end-to-end delay
compared with voice communication, but processes a set of key performance indicators (KPIs) to evaluate
users' quality of experience (QoE).
According to \cite{dobrian:sigcomm2011}, there exist five major industrial metrics for Internet streaming, in which
\emph{buffer starvation} is the most prominent.
During the playback, if the buffer turns empty, which we
call \emph{buffer starvation}, the video player goes to a buffering state and the
user sees frozen images. The measurement study in \cite{dobrian:sigcomm2011} shows that nearly
80\% of users choose to terminate the video playback before the end when 10\% of time is wasted on the buffer starvation.
Hence, it is of great importance to guarantee smooth video playback.

Owing to the significance of unraveling buffer starvation behavior, there have been a plethora of studies that aim at modeling
the delicate tradeoff between initial start-up delay and buffer starvation \cite{Liang:multimedia2008}\cite{Luan:multimedia2010}. They mainly vary in two aspects; one is the
bandwidth variation model, the other is the mathematical approach. Authors in \cite{Liang:multimedia2008} computed the starvation probability for
variable bit-rate (VBR) streaming in an extended Gilbert channel using basic probability techniques. A G/G/1 queue is considered in
\cite{Luan:multimedia2010} where the arrival and service rates are characterised by their first two moments using Brownian motion approximation.
Authors in \cite{XU:infocom2012} propose a Ballot theorem approach and a recursive approach to compute the starvation probability explicitly for
M/M/1 and M/D/1 queues. The above work, though elegant in their methodologies, is restrictive in that only a single video
stream is considered with a fixed video length (in seconds) or size (in bits). In addition,
the network is usually predetermined as a stochastic process that governs the arrival of streaming packets.
The theme of QoE optimization is to configure the
start-up delay (in seconds) or threshold (in packets) without considering what kinds of roles the network can play.
In the literature, the power of network operators is often shielded in the modeling and optimization of video streaming QoE.

From the perspective of network operators, they are more interested in predicting the QoE of a large number of video streams
than that of each individual video stream. These video streams are pulled by users from the content providers, and each video stream
may vary in content type, popularity, traffic volume, and so on. When multiple video streams traverse
the \emph{same} bottleneck,
their QoE metrics are coupled. A more congested bottleneck results in a higher probability of starvation or more starvation events.
The network operators are used to classify services into real-time and elastic classes to guarantee corresponding quality of
service levels. However, they face the difficulty to schedule Internet video streaming service for a better overall QoE. In
particular, the video streams are originated from content providers, which can only be partially controlled by the
network operators. Our previous work \cite{xu:infocom2013} models the starvation probability and the moment generating function of
starvation events in the scenario that the bandwidth is evenly shared by competing video streams.
This is one of the earliest work that tries to resolve this difficulty. A novel analytical framework is proposed, whereas
the assumption on the exponentially distributed 
video duration impedes the applicability of the proposed models.
In this paper, we stand on the side of the network operators to deal with this challenge from three coherent angles systematically:
measuring traffic pattern, modeling the probability of buffer starvation and applying the model for resource allocation.
Our study is elicited gradually by imposing three questions as follows.

\emph{(1) What is the real traffic pattern of mobile video streaming service in today's Internet?}

Intuitively, the traffic load of video streaming service is determined by the intensity of requests and the volume of requested
videos. However, the volume of a video stream cannot be simply taken as the video duration because users may terminate the
playback before the end. In other word, the streaming volume is determined by the actual viewing time, which is a private information
known only to its content provider. To unravel this mystery, we perform an in-depth measurement study of streaming
traffic pattern from a large scale dataset. This dataset, collected from one of the largest content
providers in China from september 1 to september 14, 2014,  contains nearly 200 million view records on each day.
We extract two important information that reflects the realistic streaming traffic load, i.e. the actual viewing time in every
effective record and the number of requests per hour.

\underline{Our first observation} is that the streaming traffic pattern is time-dependent in a day, and
two peaks happen at noon and in the evening. Similar request pattern exhibits periodicity on the same day in
different weeks. Hence, history traffic pattern can well predict the future traffic pattern.
\underline{Our second observation} is that the viewing time follows a hyper-exponential distribution with two time-scale separation.
This implies that there are two classes of video streams: one class has a short viewing time exponentially distributed with the
mean 94 seconds, the other class has a long viewing time exponentially distributed with the mean 1143 seconds.
The R-square test validates that our hyper-exponential distribution model is more accurate than the exponential as well as
Pareto distribution models.

\emph{(2) How can the QoE metrics be modeled?}

Our measurement study sheds light on the new design space to improve streaming QoE in cellular networks.
On one hand, the periodicity and variability of streaming traffic pattern can be utilized to design time-dependent resource
allocation strategies based on the history information.
On the other hand, the network operator may treat short and long views with different priorities so that the overall streaming
QoE can be enhanced without incurring complicated algorithm design or notable cost.
To orchestrate the QoE-oriented scheduling strategies, we need theoretical models to capture the buffer starvation probability
for video streaming services.

We build a closed-form model on starvation behavior in a bandwidth sharing wireless network. With the assumption of
geographical homogeneity in user behaviors, we map our observations to users collocating in the same wireless cell. This facilitates us to
examine the per-flow QoE metrics in a microscopic perspective. The hyper-exponential distribution of viewing time enables
us to differentiate video streams into short and long flows. Apart from the viewing time, the throughput of a video stream
is also determined by the scheduling algorithm at the base station. In general, the weighted proportional fair and
weighted round robin schedulers are approximated by discriminatory processor sharing (DPS) scheme. When all the
weights are one uniformly, they correspond to the egalitarian processor sharing (EPS) scheme. Hence, we model the
number of flows coexisting in the base station as a two-dimensional continuous time Markov process (CTMC). We then
construct ordinary differential equations (ODEs) on top of this CTMC to compute the starvation probability of video playback
in the closed-form.
Trace-driven simulation proves the accuracy of our models.
\underline{Our third observation} is that the progressive downloading significantly increases the starvation probability of
video streaming service.

\emph{(3) What are the potential applications of our models and how they can be utilized by network operators
with the minimum information on user behavior?}

One commonly-adopted resource allocation strategy is the priority scheduling. However,
our design rationale is fundamentally different from a large body of related work.
On one hand, we aim to leverage different QoE tradeoffs of short views and long views so as to
improve the overall QoE. On the other hand,
our design is based on data analytics of user behavior acquired from content providers.
It is known previously in \cite{Krishnan2013Video} that the short views are more sensitive to the number of starvations and
the start-up delay. Hence, a natural approach
to enhance the overall QoE is to configure a higher scheduling priority for short views.
\underline{Our fourth observation} from the models and trace-driven simulations manifests
that the starvation probability of short views can be significantly reduced
at the cost of slight increase of starvation probability in the long views.

The implementation of our models faces practical challenges. Before a video is played, the base station is unaware of
its viewing time by a user, hence is unaware whether it is a short or a long view. The statistical inference of viewing time relies on
the huge amount of view records which are proprietary to the content provider who is reluctant to share with the network
operator. This motivates our effort to pursue a practical approach to infer the type of a view with the minimum information from
the content provider. From all the view records in our dataset, we measure the degree of completion (DoC), i.e. the ratio of viewing
time over
video duration. Note that the videos are grouped into nine sets depending on their durations in seconds. We generalize the
empirical model in \cite{Chen:astudy2014} to capture the cumulative distribution functions (CDFs) of DoC for different ranges of
video duration. \underline{Our fifth observation} is that DoC decreases almost monotonically as the
video duration increases. With this empirical model, we propose a simple but accurate viewing time Bayesian inference scheme that
operates as the following. The network operation acquires DoC models from the content provider beforehand.
When a user requests a video, the network operator obtains the information regarding the video duration from the content
provider. Facilitated by the model of DoC, the network operator can perform a table look-up to obtain the
CDF of the viewing time, and hence the type of the view.

The remainder of this work is organised as follows. Section \ref{sec:measurement} measures the video traffic load and presents
a simple model. Section \ref{sec:model} formulates the mathematical model to capture QoE metrics at the flow level.
In section \ref{sec:qoemodel}, we compute the QoE metrics using the transient analysis of Markov chain.
The accuracy of proposed models
is validated by trace-driven simulations in section \ref{sec:simu}. The proposed modeling framework is further generalized to strip off
certain assumptions in section \ref{sec:extension}. Section \ref{sec:prediction} provides a Bayesian inference approach
to estimate viewing time.
Section \ref{sec:related} presents the literature review and section \ref{sec:conclusion} concludes this work.

\section{Measurement}
\label{sec:measurement}

In this section, we first describe the collected dataset. We then measure the intensity of video requests and the
distribution of viewing times. A hyper-exponential distribution model is presented to characterize the viewing time distribution.

\subsection{Dataset}

We collect data from one of the largest video streaming service providers in China
from September 1 to 14, 2014. There are more than 200 millions of view records everyday, in which most of
the views are through Android and Apple iOS mobile devices.
A view record is a set of information containing \emph{\{video id, video type,
play timestamp, stop timestamp, network type\}}.
The ``video type'' item consists of five major categories: \emph{movie, TV episode, music, cartoon} and \emph{sports}.
The ``play timestamp'' and ``stop timestamp'' indicate the beginning and end of a streaming session.
The ``network type'' item shows whether a user accesses video streaming service through WiFi or 3G. In our data set,
around 97.5\% of viewers use WiFi for that WiFi is usually free or of very low price while 3G traffic is much more expensive.
In our dataset, we also cleanse all the invalid records (i.e. viewing time equals to 0).

The purpose of our measurement is to reveal the realistic traffic pattern of video streaming service. Hence, we record the
number of requests every hour, and the actual viewing time of all the valid records.
Unlike \cite{Chen:astudy2014}, we do not measure the popularity of videos, and do not differentiate the video types either.
The underlying reason is that these meta information belongs to the content provider. The network operators can only
acquire the coarse-grained information on the streaming traffic pattern.

When we apply the insights of measurement to analyze the QoE metrics of video streaming in cellular networks,
two facts should be noted. Firstly, most of the view records are through WiFi networks.
Accessing WiFi networks is either free or charged with a very low price.
Hence, the views through WiFi networks are more efficaciously to reflect the human behavior in need of video streaming service.
Due to the high price charged by 3G operators, watching videos through 3G is still rare. However, we will show that streaming traffic over WiFi and 3G exhibits similar general patterns later. We envision in the near future when the price lowers down, streaming traffic will boost and follow the similar pattern as that in WiFi networks.
Secondly, our dataset, though collected from only one of the top-level content providers, can represent the
general user behaviour across different video streaming service providers.
One reason lies in that our dataset contains the view records of commercial content and user generated content.
The other is that the major streaming providers in China form several coalitions to purchase copyrighted content,
thus having the overlapping set of viewers.

\subsection{Intensity of Requests}

\begin{figure}[!tb]
\centering
\includegraphics[width=2.7in]{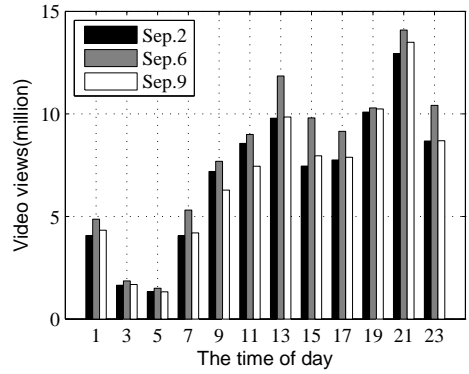}
\caption{The amount of video requests changeing over time (three days)}
\label{fig:intensity}
\vspace{-0.3cm}
\end{figure}

The study of streaming QoE requires the knowledge of traffic load, which depends on the intensity of
requests and the actual viewing time. Here, we study the former one. Fig.\ref{fig:intensity} shows the number of requests per
hour in three different days. We can see that the intensity of requests changes over time on a
daily basis. Two peaks happen at noon and in the evening, which is easy to explain: users watch more videos
during rest hours and less during working or sleeping hours. We further compare the intensity of requests
on different days and observe that similar patterns hold on different weeks.
When the traffic pattern of large-scale measurement is mapped to a concrete cellular network,
the intensity of requests in a cell also varies over time on each day, and can be well estimated
using history information. Hence, with the knowledge of request intensity, the base station can
adopt different scheduling algorithms at different time on each day, which may significantly
improve streaming QoE.

\subsection{Viewing Time Distribution}
\begin{figure*}[!tb]
\begin{minipage}[t]{0.33\linewidth}
\centering
\includegraphics[width=2.6in]{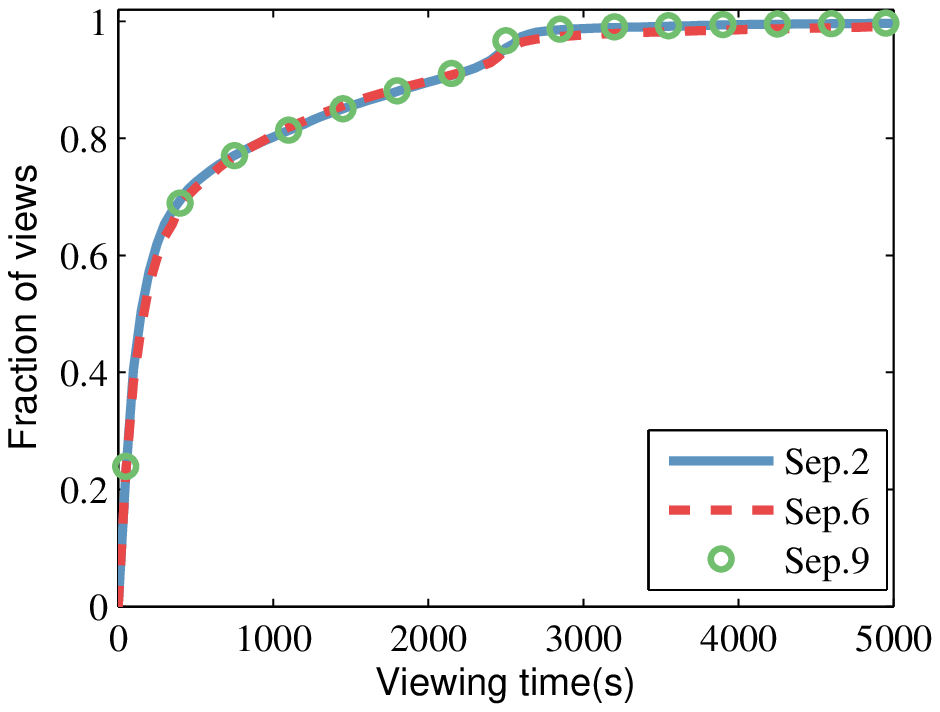}
\caption{CDF of viewing time in three different days}
\label{fig:cdf_viewtime}
\end{minipage}
\hspace{1ex}
\begin{minipage}[t]{0.33\linewidth}
\centering
\includegraphics[width=2.6in]{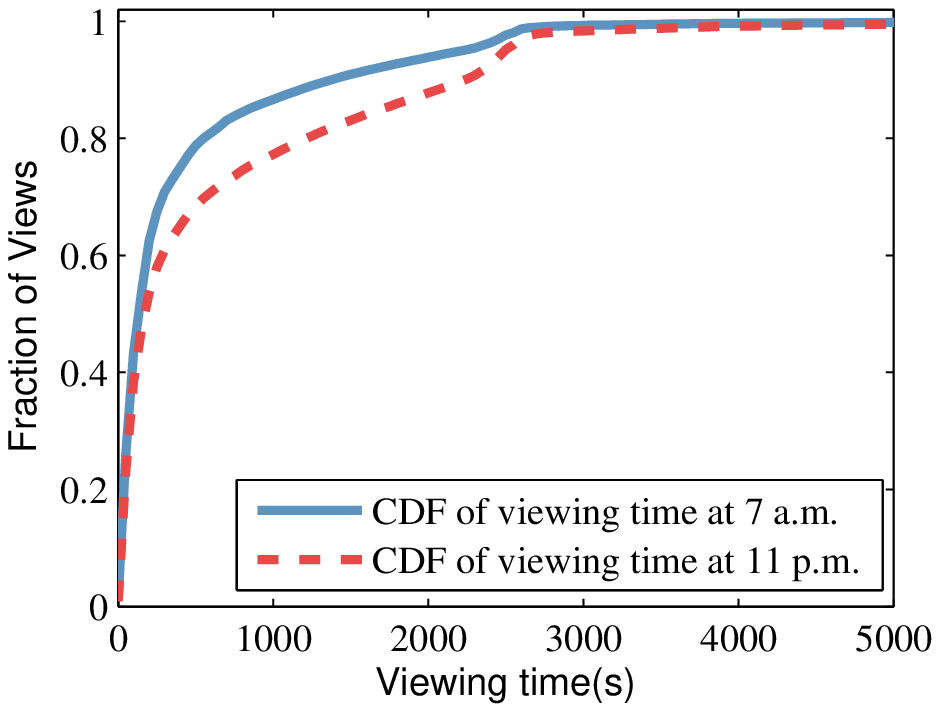}
\caption{Viewing time distribution at 7 a.m. and 11 p.m.}
\label{fig:cdf_viewtime_hours}
\end{minipage}
\hspace{1ex}
\begin{minipage}[t]{0.33\linewidth}
\centering
\includegraphics[width=2.6in]{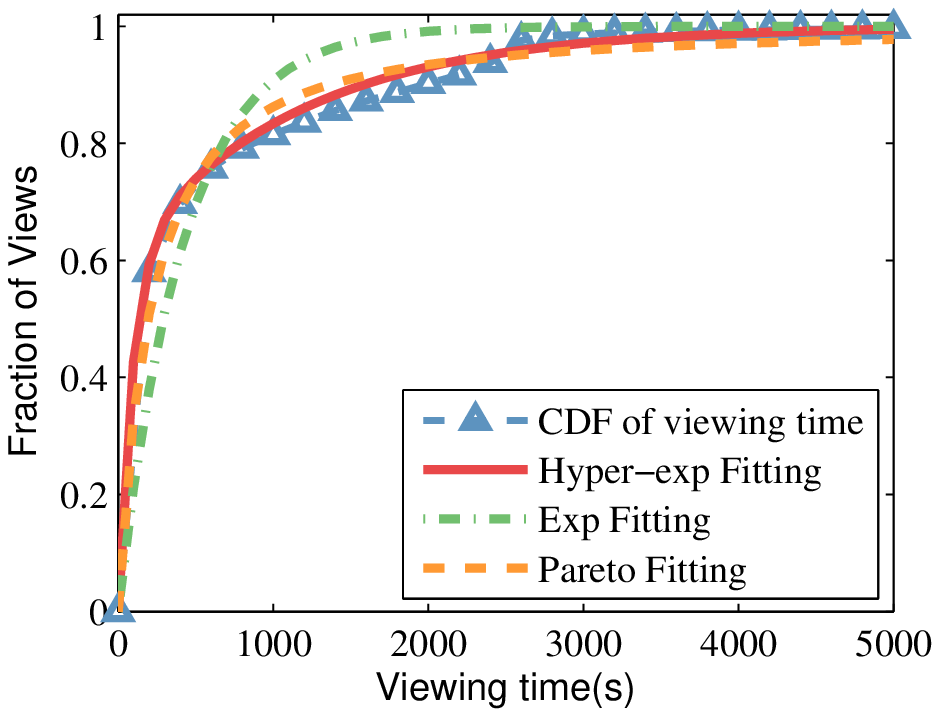}
\caption{CDF and its fitting using hyper-exponential, exponential, pareto models}
\label{fig:cdf_viewtime_fitting}
\end{minipage}
\vspace{-0.3cm}
\end{figure*}

Understanding the viewing time distribution is also crucial to model QoE metrics. To this end, we study the viewing time of all valid view records.
Fig.\ref{fig:cdf_viewtime} shows the cumulative distribution function (CDF) of the viewing time in three different days. An interesting finding is that a large fraction, approximately 70\%, of the videos have relatively short viewing time (within 500 seconds).
Meanwhile, there is a sudden increase around 2600 seconds. One reasonable explanation is that 2500 seconds
(42 mins equivalently) is the common length of a TV episode, which is an important video type.
And above all, the
viewing time distributions in different days are almost the same, which again demonstrates that history
information can be utilized to estimate the current traffic load.
We also notice that the viewing time distribution changes over time on the same day. As Fig.\ref{fig:cdf_viewtime_hours} shows, people
watch longer videos at 11 p.m. than at 7 a.m. This is in tune with reality, since people are more likely to watch long videos, such as movies,
at night, while watching short videos, such as news, sport video clips, in the morning.

The viewing time distribution has drawn a lot of attention recently. Dobrian et al
\cite{dobrian:sigcomm2011} measured the relative viewing time of users, and Xu et al
\cite{xu:infocom2013} assumed the exponentially distributed viewing time of streaming users.
However, Fig.\ref{fig:cdf_viewtime_hours} shows that the actual viewing time can hardly be approximated by the
exponential distribution. It is well known that any positive-valued distribution can be
approximated by a phase-type distribution that is the convolution of exponential distributions.
The phase-type distribution retains the memoryless property of random variables.
More phases provide better approximation while making the distribution less informative in term of
physical meaning.  We hereby propose a hyper-exponential distribution (i.e. two phases)
to fit the viewing time.

The probability density function (p.d.f.) of a hyper-exponential distribution is
\begin{equation}
f(t;p_1,p_2,\theta_1,\theta_2)=p_1\theta_1e^{-\theta_1t}+p_2\theta_2e^{-\theta_2t}
\end{equation}
where \begin{math} p_1{+}p_2{=}1 \end{math}.  Hyper-exponential distribution can be considered as the combination of two exponential distributions. The physical interpretation is that there is probability \begin{math} p_1 \end{math} of viewing time taking the exponential distribution with mean \begin{math} {1/\theta}_1 \end{math}, while there is probability \begin{math} p_2 \end{math} of viewing time taking
the exponential distribution with mean \begin{math} {1/\theta}_2 \end{math}.

Once the model is chosen, we estimate the parameters \begin{math} p_1,p_2, \theta_1 \end{math} and  \begin{math} \theta_2  \end{math}. Normally, \emph{Maximum Likelihood Estimation} (MLE) method is used because of its desirable statistical properties\cite{YoumnaBorghol2011Characterizing}. With the observed dataset (\begin{math}t_1,t_2,....,t_N \end{math}), the likelihood function can be given as (\begin{math}p_2\end{math} is removed from the parameter set since it can be obtained by \begin{math}1-p_1\end{math})
\begin{equation}
L(p_1,\theta_1,\theta_2)=\prod\nolimits_{i=1}^{N} f(t_i;p_1,\theta_1,\theta_2)
\end{equation}

The estimator of \begin{math}p_1 \end{math} can be derived by taking the partial derivative of the likelihood function w.r.t. \begin{math} p_1 \end{math}. Let it be 0,
\begin{equation}
\partial L(p_1,\theta_1,\theta_2)/\partial p_1=0.
\label{eq:MLE}
\end{equation}

The estimators of \begin{math} \theta_1 \end{math} and \begin{math} \theta_2 \end{math} can be obtained in the same way. However, the closed-form solution to Eq.\eqref{eq:MLE} is intractable because of the complicated likelihood function. Hence, we use Newton-Raphson algorithm to solve the resulting equations iteratively.

We further compare the fitting results of exponential, hyper-exponential and generalized Pareto distribution models. The p.d.f.
of exponential distribution is
 \begin{equation}
 f_{exp}(t;\theta)=\theta e^{-\theta t}
 \end{equation}
and that of generalized Pareto distribution is
  \begin{equation}
 f_{gp}(t;\xi , \sigma)=\left\{
\begin{array}{rcl}
\frac{1}{\sigma}(1+\frac{\xi}{\sigma}t)^{-(1+\frac{1}{\xi})}       &      & {\xi \ne 0}\\
\frac{1}{\sigma}e^{-\frac{t}{\sigma}}  \quad \quad \quad &      & {\xi = 0}
\end{array} \right.
 \end{equation}

The corresponding estimation results of parameters for exponential and Pareto models can also be
obtained by MLE method. Table \uppercase\expandafter{\romannumeral1} presents the parameter estimation
results and Fig.\ref{fig:cdf_viewtime_fitting} illustrates the CDF of the viewing time and three fitting curves.

We apply \emph{coefficient of determination} to validate the goodness of fit of these three models and determine which distribution best
fits our data. The coefficient of determination, also known as r-square value ($R^2$), indicates how well a statistical model matches
the dataset. It is calculated by
$$
R^2 = 1-\frac{RSS}{TSS}
$$
where $RSS$ is the residual sum of squares and $TSS$ is the total sum of squares. So, $R^2$ takes values between 0 and 1, with 0 denoting that model does not explain any variation and 1 denoting that it perfectly explains the observed variation. However, the model's Degree of Freedom will affect $R^2$. To avoid this effect, the adjusted r-square can be expressed as
$$
\overline R^2 = 1-\frac{RSS/{DoF}_{error}}{TSS/{DoF}_{total}}
$$

From the adjusted r-square values of three models in Table \uppercase\expandafter{\romannumeral2}, we can see that the hyper-exponential model fits video viewing time most accurately.
Since the viewing time can be described by hyper-exponential distribution, we develop a two-class characterization of videos.
The requested videos can be divided into two class according to their viewing time.
The videos from class-1 have exponentially distributed viewing time with mean 94s (short videos), while the videos from class-2 have exponentially distributed viewing time with mean 1143s (long videos).  This two-class characterization can help us model the flow dynamics as a two-dimensional Markov Chain.

\begin{table}[h]
\centering
\begin{tabular}{ccc|cc|c}
\hline
\multicolumn{3}{c|}{\textbf{Hyper-exp Fit} }& \multicolumn{2}{c|}{\textbf{GP Fit}} & \multicolumn{1}{c}{\textbf{Exp Fit}} \\
 $p_1$ &   $ 1/\theta_1 $    &   $1/\theta_2$   & $  \xi$  &  $ \sigma$    &  $1/\theta$                                                                                \\ \hline
 0.6011 & 94 &  1143  &    0.7823  & 210.3 &  420                                                                                     \\ \hline
 \vspace{-1em}
\end{tabular}
\caption{Parameter estimation results using MLE method}
\vspace{-0.5cm}
\end{table}

\begin{table}[h]
\centering
\begin{tabular}{l|c|c|c}
\hline
   &\textbf{Hyper-exp Fit} & \textbf{GP Fit} &\textbf{ Exp Fit}\\ \hline
\textbf{ Adjusted r-square} & 0.9958 & 0.9753 & 0.9032 \\  \hline
\end{tabular}
\caption{Adjusted r-square values of three models}
\vspace{-0.5cm}
\end{table}

\begin{figure}[!tb]
\centering
\includegraphics[width=2.7in]{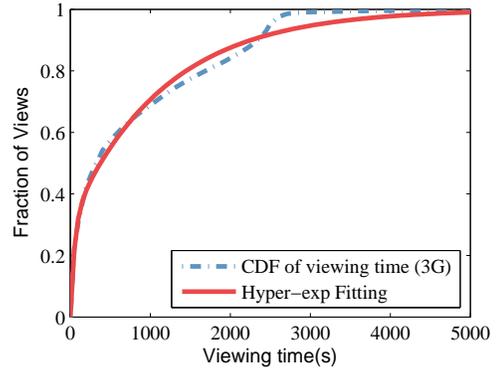}
\caption{CDF of viewing time (3G) and its hyper-exponential fitting}
\label{fig:cdf_viewtime_3G}
\vspace{-0.3cm}
\end{figure}

Although the above measurements exploit view records in both WiFi and 3G networks, we show in Fig.\ref{fig:cdf_viewtime_3G} that
streaming traffic over 3G exhibits similar general patterns. The viewing time of 3G video users fits the hyper-exponential distribution
very well (with different parameters).

\section{Problem Description and System Model}
\label{sec:model}

In this section, we define the QoE metrics for video streaming service and present the basic queueing model of playout buffer.


\subsection{QoE Metrics}

The QoE metrics are usually defined in a variety of ways. In the seminal work \cite{dobrian:sigcomm2011},
the authors enumerate five industry-standard video quality metrics: \emph{join time}, \emph{buffering ratio}, \emph{rate of buffering events}, \emph{average bitrate} and \emph{rendering quality}.
Among them, the buffering ratio and rate of buffering events have the largest impact on the user engagement \cite{dobrian:sigcomm2011} that indirectly reflects the subjective human perception.
The buffering ratio represents the fraction of total time wasted in buffer starvation. It is mainly determined by the
throughput process for a fixed viewing time.
The rate of buffering events captures the frequency of induced starvations observed by the user.
It involves both the start-up delay and rebuffering delay that are usually configured by the content provider.
The buffering ratio and rate of buffering events are positively correlated so that a large buffering ratio may result in
more starvation events. These two metrics are proposed for the \emph{ex post} measurement
study on the buffer starvation, while not frequently used for QoE modeling in the literature.
They are defined for engineering convenience and they are coupled with the parameters such as start-up and rebuffering
delays, which is extremely difficult to be analyzed.
Besides, the joining time is usually configured by the content providers in the form of advertising duration.
The average bit-rate and rendering quality are pertinent to content itself. Hence, we only consider the QoE metrics about
buffer starvation throughout this work. For clarity, we redefine two QoE metrics rigidly to capture the severity
of buffer starvation.

\begin{itemize}

\item \textbf{Starvation Probability.} It is a commonly adopted metric (e.g. in \cite{A:interruptions2011}\cite{Luan:multimedia2010}) that indicates the probability of meeting buffer starvation during the video playback.

\item \textbf{Mean DT/VT Ratio.} Here, DT/VT refers to the ratio of downloading time over viewing time. 
The mean DT/VT ratio serves as the lower bound of buffering ratio since the latter is extremely complicated and has
not been modeled so far.

\end{itemize}

Once we derive the probability distributions of starvation event and DT/VT ratio, other QoE metrics such as the distribution of 
the number of starvation events can be obtained accordingly. To the best of our knowledge, 
this is the first time that a QoE metric related to buffering ratio is considered for modeling.
We also need to keep in mind that the starvation probabilities (resp. the mean DT/VT ratio)
may be different among the short and long views.

\subsection{Network Model and Resource Allocation}

We consider a wireless data network that supports a number of flows. When a new flow ``joins'' the network, it requests the
streaming service from a media server. After the connection has been built, the streaming packets are transmitted through the base
station (BS). The streaming flows have \emph{finite sizes}, which means that a flow ``leaves'' the network once the
transmission completes.
Note that each active user cannot watch more than one streams at the mobile device simultaneously. Hence, we use the terms
``flow'' and ``user'' interchangeably. The flows are competing for finite capacity (or resource). When the number of flows sharing the
bottleneck increases, network congestion occurs. The decrease of per-flow throughput may result in the undesirable playback
interruption of video streaming services. The arrival and departure of flows further cause throughput fluctuation of the concurrent
flows. To summarize, with the dynamics of co-existing flows in the bottleneck, per-flow throughput is not a random variable that has
been studied in the literature, but a continuous time stochastic process.

In wireless data networks, a streaming flow may traverse both wired and wireless links, whereas the BS is the bottleneck for the
sake of limited channel capacity. Most of Internet streaming providers such as Youtube and Youku use TCP/HTTP protocols to
deliver streaming packets. TCP congestion control scheme tries to send as more packets as possible to the user (by consistently
increasing the congestion window) in order to explore the available bandwidth. The sender reduces the congestion window when
packet drops have been detected. The packet drops happen when the number of backlogged packets exceeds a certain threshold,
e.g. the maximum queue length in the DropTail and the minimum threshold in the random early detection (RED) active queue
management scheme. Therefore, it is reasonable to assume that the queue of an active flow is always backlogged at the BS. The
packet losses rarely happen under adversary wireless channel conditions. The reason lines in that the adaptive coding and
modulation in the physical layer, and ARQ scheme at the MAC layer can effectively avoid TCP packet loss caused by channel variation. A recent measurement study validates that TCP packet loss rate is usually less than 0.1\% in 3G/4G channels \cite{imc2013:chen}.

Our focus is to investigate the impact of resource allocation strategy at the BS on the starvation metrics.
According to the large-scale measurement study, we are aware that video requests can be partitioned into two classes, the short
and the long views. In each class, the viewing time of video requests is exponentially distributed.
When the large-scale video requests are mapped to a cell, it is reasonable to postulate that some of the requests
are short views and some others are long views. The network operator faces an important question whether he will offer a
higher priority to the short views or the long views so to improve the overall QoE. We hereby introduce a generalized
scheduling mechanism, namely Discriminatory Processor Sharing (DPS), which performs bandwidth allocation at the flow
level with priorities. An example is used to highlight the basic principle of DPS. Suppose that the BS is shared by
$n_1$ video streams in the short class and $n_2$ video streams in the long class. The weights $\phi_1$ and $\phi_2$ are
endowed to the short class and long class respectively. Then, every class-$k$ flow obtains the fraction of total throughput by
$\frac{\phi_1}{\phi_1n_1 + \phi_2n_2}$ for $k=1, 2$. If $\phi_1 > \phi_2$, a video stream in class-1 obtains a higher
throughput than the one in class-2, and vice versa. The range of DPS applications is broad, where the most natural one is
to model weighted Round-Robin (WRR) scheduling. The BS in cellular networks performs resource allocation for each flow at
every time slot such that WRR scheduling can be easily implemented.

\subsection{Basic Queueing Model of Playout Buffer}

We consider a wireless cellular network that supports up to $K$ simultaneous flows. With admission control, the
overloading of the BS can be regulated. We make the following assumptions.

\begin{itemize}

\item \textbf{Uniformity of User Behavior.} We assume that the user behavior in a cell is consistent with the overall
user behavior. Hence, the viewing time in a cell follows the same hyper-exponential distribution, and is classified into the
short type and long type.

\item \textbf{Continuous time playback.} The service of video contents is regarded as a continuous process, instead of a
discrete rendering of adjacent video frames spaced by a fixed interval. This assumption is commonly used in the
literature (see \cite{TOMCCAP04:wang}). We assume that all the flows have identical video bit-rate.

\end{itemize}

The assumption of uniformity of user behavior may not be always accurate. For instance, the user behavior in a residential
region can be different from that in a commercial region. To acquire the exact user behavior, the network operator needs to
store all the streaming packets in each cell and performs deep packet inspection (DPI). However, this is usually not practical
due to two reasons. On one hand, the storage and computation costs are prohibitively expensive.
On the other hand, the viewing time and video duration are unknown to the network operator. The network operator can hardly do
anything toward the on-going video streams.

In line with the large-scale measurement, the viewing time in
each class $k$ is exponentially distributed with the mean $1/\theta_k$, $(k=1, 2)$.
Let $C$ be the capacity of the wireless channel in bits per-second (bps) and $Bitrate$ be the playback speed of video
streams in bits per-second. Then, the volume (in bits) of a class-$k$ video stream is exponentially distributed with the mean
$\frac{Bitrate}{\theta_k}$.
We denote by $\lambda_k$ the Poisson arrival rate of new video streams belonging to class $k$, $(k=1, 2)$,
where the total arrival rate
is computed by $\lambda = \lambda_1 + \lambda_2$. Let $p_k$ be the fraction of class $k$ flows, i.e. $p_k = \frac{\lambda_k}{\lambda}$. Then, the dynamics of coexisting flows in the cell can be depicted as a two-dimensional
continuous time Markov chain (CTMC) with a finite state space: one dimensional is the number of short views,
and the other is the that of long views. The throughput of a video stream depends on the Markov state of the BS.

The development of the QoE models relies on the so-called ``tagged flow'' approach, in which one keeps track of the system evolution
from the arrival until the departure of a tagged video stream. At any time $t$, the tagged flow sees $i$ other
short flows and $j$ other long flows in a finite space $S:=\{(i,j)| i+j \leq K{-}1, i\geq 0, j\geq 0\}.$ The sum of $i$ and $j$ is
no larger than $K{-}1$ simply because the tagged flow is already admitted by the BS.
The video playback consists of the initial start-up/rebuffering phase and the playback phase.
Denote by $Q(t)$ the length of playout buffer at time $t$ in the client side.
Here, $Q(t)$ is not measured in packets or bits that are usually adopted
in queueing theory, but is rescaled in the seconds of video content that has not been played.
In the initial start-up/rebuffering phase, the media player of a viewer keeps downloading video content without playing it.
At each state $(i,j)$, the downloading speed of the class-$k$ tagged flow is
\begin{eqnarray}
b_{i,j}^k := \frac{C\phi_k}{Bitrate\cdot \big((i+ 1_{\{k=1\}})\phi_1 + (j+1_{\{k=2\}})\phi_2\big)}.
\end{eqnarray}
The indicator $1_{\{k=1\}}$ (resp. $1_{\{k=2\}}$) is 1 if $k$ equals to 1 (resp. 2) and 0 otherwise.
This means that at state $(i,j)$ the rate of playout buffer increment is $b_{ij}^k$ for a class-$k$ tagged flow.
While in the playback phase, the viewer downloads and plays at the same time. Thus, at each state $(i,j)$, the
rate of playout buffer increment for a class-$k$ tagged flow, denoted by $c_{i,j}^k$, is given by
$c_{i,j}^k := b_{i,j}^k - 1$. One can observe that the dynamics of playout buffer length also depends on the DPS weights
$\phi_1$ and $\phi_2$. In what follows, we will model the QoE metrics of the tagged flow, and investigate the impact of
DPS strategy on the QoE metrics.

\section{Analysis of QoE Metrics}
\label{sec:qoemodel}
In this section, a set of Markov chains are constructed to characterize the dynamics of streaming flows at the BS.
We further present closed-form solutions to the starvation probability and the DT/VT ratio.

\subsection{Basic Markov Models of Flow Dynamics}

We adopt the ``tagged flow'' approach to analyze the QoE metrics. Therefore, it is necessary to examine all the possible phases
in the service of a video stream. We construct three Markov processes to characterize the flow-level dynamics at different
stages: before the arrival of the tagged flow (denoted as \textbf{MC1}), the start-up phase of the tagged flow
(denoted as \textbf{MC2}) and the playback phase of the tagged flow (denoted as \textbf{MC3}). If not mentioned explicitly, we
suppose the tagged flow belongs to class-1. All the analyses can be reproduced directly if the tagged flow belong to class-2.

\begin{figure}[h]
\centering
\includegraphics[width=3.3in]{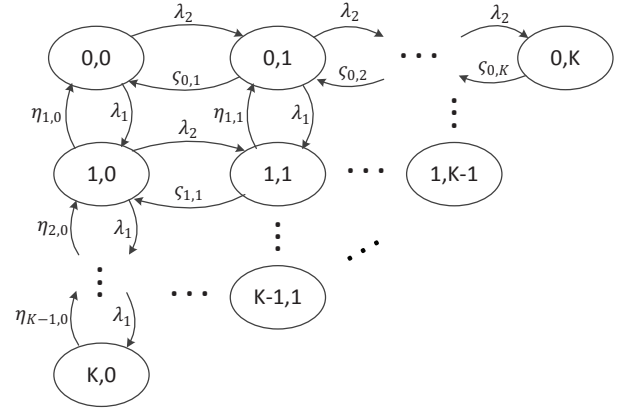}
\vspace{-0.1cm}
\caption{Markov chain for flow dynamics before arrival of tagged flow}
\label{fig:mc1}
\end{figure}

We have two purposes when constructing \textbf{MC1}. One is to calculate the probability of rejecting a streaming request
due to the admission control policy, the other is to obtain the stationary distribution of system states at an arbitrary time that
the tagged flow joins. The class-$k$ video requests arrive with a Poisson rate $\lambda_k$ for $k\in\{1,2\}$, in which the total
arrival rate is $\lambda=\lambda_1+ \lambda_2$. The service time of a video stream at a state is equal to the
size of video request in bits divided by its throughput. Because the viewing time follows the hyper-exponential distribution,
the service time of each class is also exponentially distributed. Therefore, the dynamics of the number of coexisting flows can be
modeled as a two-state continuous time Markov process shown in Fig.\ref{fig:mc1}. The state space of \textbf{MC1} is
$S\cup\{i,j| i\geq 0, j\geq 0, i+j=K\}$ because the tagged flow has not arrived yet so to have at most $K$ coexisting flows.
The transition rate from state $(i,j)$ to state $(i+1,j)$ is $\lambda_1$ and from state $(i,j)$ to state $(i, j+1)$ is $\lambda_2$.
Let $\eta_{i,j}^1$ be the transition rate from state $(i,j)$ to state $(i-1,j)$ and $\eta_{i,j}^2$ be that from state
$(i,j)$ to $(i, j-1)$.
Here, $\eta_{i,j}^k$ is the product of the number of class-$k$ flows and the inverse of service time of a class-$k$ flow
according to the definition of Markov state transition.
Hence, there have $\eta_{i,j}^1 = \frac{i\phi_1\theta_1C}{Bitrate\cdot (i\phi_1+j\phi_2)}$ and
$\eta_{i,j}^2 = \frac{j\phi_2\theta_2C}{Bitrate\cdot (i\phi_1+j\phi_2)}$. Denote by $z_{i,j}$ the stationary probability that
there are $i$ class-1 flows and $j$ class-2 flows coexisting at the BS. Here, $z_{i,j}$ can be easily computed from the standard
Markov chain analysis, in which the detailed computation is omitted here. When the tagged flow arrives, it observes
the system at state $(i,j)$ with the probability $z_{i,j}$. Due to the admission control policy, it is rejected with the probability
$p_{rej} = \sum_{i=0, j=K-i}^{K} z_{i,j}$. If the tagged flow is accepted, it enters the start-up phase. We want to compute
the conditional probability $\pi_{i,j}$ that the tagged flow observes $i$ other class-1 flows and $j$ other class-2 flows after
being accepted. It is easy to yield
\begin{eqnarray}
\pi_{i,j} = \frac{z_{i,j}}{1 - p_{rej}}, \quad (i,j)\in S.
\end{eqnarray}
So far, we have derived the stationary distribution that the tagged flow observes the system at state $(i,j)$ upon its arrival if
it is accepted.

\begin{figure}[h]
\centering
\includegraphics[width=3.3in]{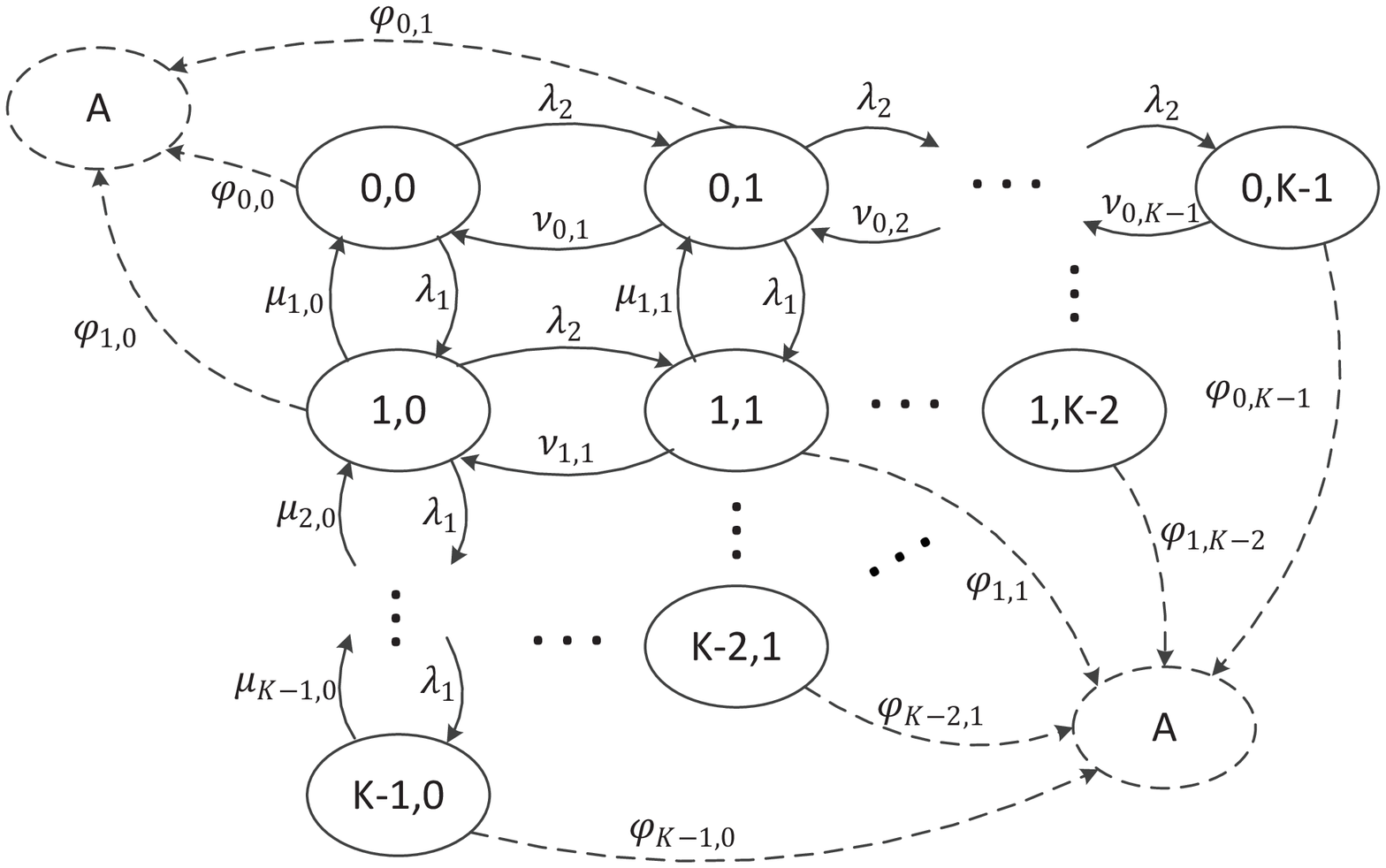}
\caption{Markov chain for flow dynamics observed by tagged flow in start-up phase and playback phases}
\label{fig:mc2}
\end{figure}

Our purpose of building \textbf{MC2} is to analyze the state distribution when the initial prefetching phase ends. As is well known,
the content provider usually configures a fixed start-up threshold either for reducing starvations or for displaying advertisements.
For analytical convenience, we define the start-up threshold not as an absolute time duration, but as the prefetched video
content measured in seconds (e.g. ten seconds of video content). In \textbf{MC2}, each state $(i,j)$ refers to that
the tagged flow observes $i$ other class-1 flows and $j$ other class-2 flows at the start-up phase. Since the tagged flow always
exists, the total number of other flows is no more than $K-1$ such that the state space of \textbf{MC2} is $S$.
We plot \textbf{MC2} in Fig.\ref{fig:mc2} in which the state $\mathbf{A}$ is excluded (it will be used in \textbf{MC3} later on).
Denote by $\mu_{i,j}$ the transition rate from state $(i,j)$ to state $(i-1,j)$, and by $\nu_{i,j}$ the transition rate from
state $(i,j)$ to state $(i,j-1)$. To compute $\mu_{i,j}$ and $\nu_{i,j}$, we first assume that the tagged flow belongs to class-1.
At state $(i,j)$ of \textbf{MC2}, the throughput of a class-1 flow is $\frac{\phi_1C}{(i+1)\phi_1 + j\phi_2}$ in bps, and that of
a class-2 flow is $\frac{\phi_2C}{(i+1)\phi_1 + j\phi_2}$ in bps. The service
time of a class-1 flow is obtained by $\frac{\phi_1\theta_1 C}{Bitrate\cdot ((i+1)\phi_1 + j\phi_2)}$, and that of a class-2 flow is
$\frac{\phi_2\theta_2 C}{Bitrate\cdot ((i+1)\phi_1 + j\phi_2)}$. Hence, at state $(i,j)$, the
transition rate to state $(i-1,j)$ is $\mu_{i,j} := \frac{i\phi_1\theta_1 C}{Bitrate\cdot ((i+1)\phi_1 + j\phi_2)}$, and
the transition rate to state $(i, j-1)$ is $\nu_{i,j}:= \frac{j\phi_2\theta_2 C}{Bitrate\cdot ((i+1)\phi_1 + j\phi_2)}$.
Alternatively, if the tagged flow is a class-2 flow, the transition rate can be analyzed in the same way.

We construct \textbf{MC3} to analyze the starvation probability of the tagged flow, given the state $(i,j)$
and the playout buffer length $q$ at the beginning of the playback phase. \textbf{MC3} is shown in Fig.\ref{fig:mc2} where
$\mathbf{A}$ is an absorbing state. The absorbing state $\mathbf{A}$ means that the tagged flow finishes its downloading and
leaves the system. The throughput of the tagged flow is $\frac{\phi_1\theta_1 C}{(i+1)\phi_1 + j\phi_2}$ so that the
transition rate from state $(i,j)$ to $\mathbf{A}$ can be solved by
$\varphi_{i,j}:=\frac{\phi_1\theta_1 C}{Bitrate\cdot((i+1)\phi_1 + j\phi_2)} = \frac{\mu_{i,j}}{i}$.
Note that there is only one absorbing state in Fig.\ref{fig:mc2}. We plot two $\mathbf{A}$ circles in order to
avoid too many intersections of the state transition curves to the absorbing state $\mathbf{A}$.

\subsection{Analysis of Starvation Probability}
Obviously, starvation only happens during playout phase. Let \(c_{i,j}:=b_{i,j}-1\) denote the increasing rate of buffer length during playout
phase and \(W_{i,j}(q)\) denote the starvation probability with playout buffer length \(q\) at state $(i,j)$. Notice that at the absorbing
state $A$, the tagged flow finishes video buffering. Thus, \(W_A(q)=0\) for any \(q\ge 0\). Considering an infinitesimal
time interval \([0,h]\), there are
seven possible events when the buffer length changes from \(q\) to \(q+c_{i,j}h\):
\begin{itemize}
\item arrival of one class-1 flow;
\item arrival of one class-2 flow;
\item departure of one class-1 flow (not the tagged flow);
\item departure of one class-2 flow;
\item departure of tagged flow;
\item occurrence of at least two events;
\item no changes.
\end{itemize}
Then, we can derive a set of equations for \(W_{i,j}(q)\)
\begin{align}
W_{i,j}(q)&=(1-(\lambda+\mu_{i,j}+\varphi_{i,j}+\nu_{i,j})h)W_{i,j}(q+c_{i,j}h)\nonumber \\
&+\lambda_1hW_{i+1,j}(q+c_{i,j}h)+\lambda_2hW_{i,j+1}(q+c_{i,j}h) \nonumber\\
&+\mu_{i,j}hW_{i-1,j}(q+c_{i,j}h)+\nu_{i,j}hW_{i,j-1}(q+c_{i,j}h) \nonumber \\
&+\varphi_{i,j}hW_A(q+c_{i,j}h)+o(h).
\label{eq:ode_w_no1}
\end{align}
With \(W_A(q+c_{i,j}h)=0\), Eq.\eqref{eq:ode_w_no1} can be expressed as
\begin{flalign}
&\frac{1}{h}(W_{i,j}(q)-W_{i,j}(q+c_{i,j}h))=\nu_{i,j}W_{i,j-1}(q+c_{i,j}h)\nonumber\\
&-(\lambda_1+\lambda_2+\mu_{i,j}+\varphi_{i,j}+\nu_{i,j}))W_{i,j}(q+c_{i,j}h) \nonumber\\
&+\lambda_1W_{i+1,j}(q+c_{i,j}h)+\lambda_2W_{i,j+1}(q+c_{i,j}h) \nonumber\\
&+\mu_{i,j}W_{i-1,j}(q+c_{i,j}h)+o(h)/h.
\label{eq:ode_w_no2}
\end{flalign}
When \(h\to0\), the left side of Eq.\eqref{eq:ode_w_no2} is the ordinary differential of \(W_{i,j}(q)\) over \(q\). With \(W_A(q+c_{i,j}h)=0\), we can derive a set of ordinary differential equations (ODEs)
\begin{flalign}
c_{i,j}\dot W_{i,j}&=(\lambda_1+\lambda_2+\mu_{i,j}+\varphi_{i,j}+\nu_{i,j}))W_{i,j}(q)\nonumber\\
                   &-\lambda_1W_{i+1,j}(q)-\lambda_2W_{i,j+1}(q)-\mu_{i,j}W_{i-1,j}(q) \nonumber\\
                   &-\nu_{i,j}W_{i,j-1}(q).
\end{flalign}
The above equations can be rewritten in a matrix form as
\begin{equation}
\dot {\mathbf W}(q)=\mathbf M_W \mathbf W(q)
\label{eq:ode_w_no4}
\end{equation}
where \({\mathbf W(q)=(W_{i,j}(q))}^T\) is a vector with indices \(l(i,j)=\sum_{m=0}^{i-1}{(K-1-m)}+j+1\) for all \((i,j)\) that has
\(i\le K-1\) and \(j\le K-1-i\). The matrix $\mathbf M_W$ is
\begin{equation}
\left(\begin{array}{ccccc}
    \mathbf S_0 & \mathbf L_0 & & & \\
    \mathbf U_1 & \mathbf S_1 & \mathbf L_1 & & \\
        &\ddots&\ddots&\ddots& \\
        &       &\mathbf U_{N_1(0)-1}&\mathbf S_{N_1(0)-1}&\mathbf L_{N_1(0)-1}\\
        &       &            &\mathbf U_{N_1(0)}&\mathbf S_{N_1(0)}
        \end{array}\right)
\end{equation}
where
$$
\mathbf L_i= \begin{pmatrix}\begin{smallmatrix}
         \frac{-\lambda_1}{c_{i,0}} & & \\
           &\ddots&\frac{-\lambda_1}{c_{i,N_2(i)}}\\
           0&\cdots&0
           \end{smallmatrix}\end{pmatrix},
\mathbf U_i=\begin{pmatrix}\begin{smallmatrix}
      \frac{-\mu_{i,0}}{c_{i,0}} & & &0 \\
      &\ddots& &\vdots \\
      &&\frac{-\mu_{i,N_2(i)}}{c_{i,0}}&0
     \end{smallmatrix}\end{pmatrix}
$$\\
and
$$
\mathbf S_i=\begin{pmatrix}\begin{smallmatrix}
       S_i(0) & S_i^+(0) & & & \\
    S_i^-(1) & S_i(1) & S_i^+(1) & & \\
        &\ddots&\ddots&\ddots& \\
        &       &            &S_i^-(N_1(0))&S_i(N_1(0))
      \end{smallmatrix}\end{pmatrix}
$$
with $S_i(j)= \frac{\lambda+\mu_{i,j}+\varphi_{i,j}+\nu_{i,j}}{c_{i,j}}$; $S_i^-(j)=\frac{-\nu_{i,j}}{c_{i,j}}$ and $S_i^+(j)=\frac{-\lambda_2}{c_{i,j}}$. The size of $\mathbf S_i$ is $(K-i)\times (K-i)$, the size of $\mathbf L_i$ is $(K-i+1)\times (K-i)$ and the size of $\mathbf U_i$ is $(K-i)\times (K-i+1)$.

Because $\mathbf M_W$ is a tridiagonal block matrix, $\mathbf M_W$ can be decomposed as $\mathbf M_W=\mathbf D_W\mathbf \Lambda_W \mathbf D_W^{-1}$ where $\mathbf D_W$ contains all the eigenvectors and $\mathbf{\Lambda_W}$ is a diagonal
matrix containing all the eigenvalues. Then, the solution to Eq.\eqref{eq:ode_w_no4} is
\begin{equation}
\mathbf W(q)= \mathbf D_W\exp(\mathbf \Lambda_Wq)\mathbf D_W^{-1}\cdot \mathbf W(0).
\end{equation}
It is obvious that the initial conditions satisfy $W_{i,j}(0)=1$ for all $(i,j)$ with $c_{i,j}< 0$. According to the results in
\cite{xu:infocom2013}, $W_{i,j}(0)$ for all $(i,j)$ with $c_{i,j}\ge 0$  can be obtained by solving $\overline W_{i,j}=0$ for all $(i,j)$ with $c_{i,j}\ge 0$, where $\overline {\mathbf W}:=\mathbf D_W^{-1}\cdot \mathbf W(0)$.

With the startup threshold $q_a$, the starvation probability of the tagged flow is $W_{i,j}(q_a)$ when the tagged flow
begins the video playback at state $(i,j)$. Let $\mathbf \pi$ denote the state distribution upon the tagged flow's arrival. According to \emph {Possion Arrivals See Time Average (PASTA)} discipline, the distribution of the numbers of flows when the
tagged flow arrives is the same as the Markov chain's steady-state distribution.
Let $\mathbf V(0;q_a)$ denote the state transition matrix during startup phase. The analysis of $\mathbf V(0;q_a)$ is similar to that of $\mathbf W(q_a)$, which is omitted due to lack of space (see in Appendix A).

Thus, the state distribution when playback begins can be computed by $\mathbf \pi \cdot \mathbf V(0;q_a)$. Besides, starvation only happens to flows whose viewing times are longer than $q_a$. So, the final starvation probability can be expressed as
\begin{eqnarray}
\mathbb{P}_s(q_a) &=& \mathbb{P}\{T_{video}>q_a\}\cdot \mathbf \pi \cdot \mathbf V(0;q_a)\cdot \mathbf W(q_a) \nonumber \\
 &=&\exp(-\theta_1q_a)\cdot \mathbf \pi \cdot \mathbf V(0;q_a)\cdot \mathbf W(q_a). \nonumber
\end{eqnarray}

\subsection{Analysis of Mean DT/VT Ratio}

Assuming the viewing time of the requested video is $q_v$,  we denote by $S_{i,j}(q;q_v)$ the mean downloading time of tagged flow that arrivals at state $(i,j)$ with buffered video duration $q$. In an infinitesimal interval \([0,h]\), the length of already buffered video changes from $q$ to $q+b_{i,j}h$.
Following the same technique we use in starvation probability, we obtain a set of equations for $S_{i,j}(q;q_v)$ as follows
\begin{flalign}
&b_{i,j}\dot S_{i,j}(q;q_v)=(\lambda+\mu_{i,j}+\nu_{i,j})S_{i,j}(q;q_v)-\lambda_1S_{i+1,j}(q;q_v)\nonumber\\
                   &-\lambda_2S_{i,j+1}(q;q_v)-\mu_{i,j}S_{i-1,j}(q;q_v)-\nu_{i,j}S_{i,j-1}(q;q_v)-1\nonumber
\end{flalign}
Let $S(q;q_v)$ be a matrix of tagged flow's downloading time at different states. The above equations can be rewritten as
\begin{equation}
\dot {\mathbf S}(q;q_v)=\mathbf M_V \mathbf S(q;q_v)-\{1/b_{i,j}\}
\label{eq.11}
\end{equation}
where $\{1/b_{i,j}\}$ is a column vector.
We can solve Eq.(\ref{eq.11}) directly by
\begin{flalign}
\mathbf S(q;q_v)= &-\mathbf D_V \cdot diag\{\frac{1}{\delta_V^{i,j}}(e^{\delta_V^{i,j}q}-1)\} \cdot \mathbf D_V^{-1}\cdot \{1/b_{i,j}\}   \nonumber \\
&+\mathbf D_V\exp(\mathbf \Lambda_Vq)\mathbf D_V^{-1}\cdot \mathbf S(0;q_v)
\end{flalign}

where $\delta_V^{i,j}$ is the $l(i,j)^{th}$ eigenvalue in $\Lambda_V$. When $\delta_V^{i,j}$ is 0, the term $\frac{1}{\delta_V^{i,j}}(e^{\delta_V^{i,j}q-1})$ equals 0. Obviously, $S_{i,j}(q_v;q_v)$ is 0 for all $(i,j) \in S$, then the mean sojourn time is obtained by
\begin{equation}
\mathbf S(0;q_v)= \mathbf D_V \cdot diag\{\frac{1}{\delta_V^{i,j}}(1-e^{-\delta_V^{i,j}q})\} \cdot \mathbf D_V^{-1}\cdot \{1/b_{i,j}\}
\end{equation}

Let $DV(q_v)$ denote the DT/VT ratio when the viewing time of tagged flow is $q_v$, then $DV(q_v)$ can be expressed as
\begin{equation}
\mathbf {DV}(q_v)=\frac{\mathbf S(0;q_v)}{q_v}.
\end{equation}

Notice that viewing time $q_v$ is exponential distributed with mean $1/\theta_1$, then the mean DT/VT ratio of tagged flow $\mathbb {DV}$ can be finally expressed as\begin{flalign}
\mathbb {DV}&=\pi \cdot \int^{+ \infty}_0 \mathbf {DV}(q_v)  \theta_1 e^{-\theta_1q_v} d{q_v} \nonumber \\
&=\pi \cdot \int^{+ \infty}_0  \frac{\mathbf S(0;q_v)}{q_v} \theta_1 e^{-\theta_1q_v} d{q_v} \nonumber \\
& =\pi \cdot \mathbf D_V \cdot diag\{\frac{1}{\delta_V^{i,j}} [E_i((\delta_V^{i,j}+\theta_1)\cdot 0^+) \nonumber \\
&-E_i(\theta_1 \cdot 0^+)]\} \cdot \mathbf D_V^{-1}\cdot \{1/b_{i,j}\}.
\end{flalign}

\section{Trace-driven Simulation}
\label{sec:simu}

In this section, the accuracy of our QoE models is validated through trace-driven simulation.
We further provide important insights on the design of DPS scheduler.

\subsection{Parameters and Method}

In the trace-driven simulation, video requests are randomly picked from our dataset,
while in the pure simulation, they are generated from the predetermined random distributions.
In other words, we do not need to configure the viewing times and the percentage of long or short views in the trace-driven simulation.
In general, the pure simulation is used to validate the correctness of models, while not sufficient to evaluate their accuracy.
The pure simulation asymptotically converges to the mathematic model as long as the number of stochastic events is large enough,
which motivates our trace-driven simulation.

We consider a wireless network with the capacity $C=5$Mbps. The BS can admit up to $K=10$ video streams.
The parameters for QoE models follow the measurement study in Section \ref{sec:measurement}:
an incoming flow belongs to class-1 with probability $p_1=0.6$ and belongs to class-2 with probability $p_2=0.4$;
the viewing times of class-1 and class-2 are exponentially distributed with the mean $1/\theta_1=94$s and
$1/\theta_2=1143$s respectively. The video bitrate is uniformly set to 980 Kbps.
We begin with egalitarian processor sharing queue (EPS) of the BS by setting $\phi_1=\phi_2$
that yields the average departure rates $\psi_1=\frac{C\theta_1}{Bitrate}=0.054$ and $\psi_2=0.0045$. To avoid arbitrarily choosing the arrival rate of
video requests, we set up the traffic load of the BS to be $\rho=0.96$. This implies that the BS is at a heavy load regime
while still not exceeding the capacity region. Given the expression $\lambda=\rho /(p_1/\psi_1+p_2/\psi_2)$,
the intensity of video requests is given by $\lambda = 0.0095$.

When discriminatory process sharing (DPS) scheduling is adopted at the BS, the trace-driven simulation encounters the difficulty of
differentiating which class an incoming flow belongs to. In the raw dataset, there does not exist the class property of video requests.
Therefore, to carry out DPS scheduling, we present a simple \emph{Bayesian} approach to
designate the class of every randomly selected video request. The priori knowledge includes the hyper-exponential distribution of
all the viewing times and the viewing time of incoming flows. In the interval $[t, t + \Delta t]$ with small enough $\Delta t$,
the probability that a viewing time falls in this interval is $(p_1\theta_1 e^{-\theta_1t} + p_2\theta_2 e^{-\theta_2t})\Delta t$.
We denote by $\gamma$ the probability that an incoming flow with the viewing time $t$ belongs to class-1.
Then, according to Bayesian rule, $\gamma$ is solved by
\begin{eqnarray}
\gamma = \frac{p_1\theta_1 e^{-\theta_1t}}{p_1\theta_1 e^{-\theta_1t} + p_2\theta_2 e^{-\theta_2t}}.
\label{eq:bayesian}
\end{eqnarray}
Then, we generate an evenly distributed random variable in the range $[0,1]$. If this random variable is less than $\gamma$,
the incoming flow with viewing time $t$ is regarded as a class-1 flow, and otherwise it is deemed as a class-2 flow.
After preprocessing the flow type, the BS can perform DPS scheduling by
configuring different priorities to class-1 and class-2 flows.

\subsection{QoE Metrics and Their Insights}

\textbf{Starvation Probability.}
In this set of experiments, we will evaluate the overall starvation probability, and the starvation probabilities when the
playback process begins at different states.

\begin{figure*}[!tb]
\begin{minipage}[t]{0.33\linewidth}
\centering
\includegraphics[width=2.5in]{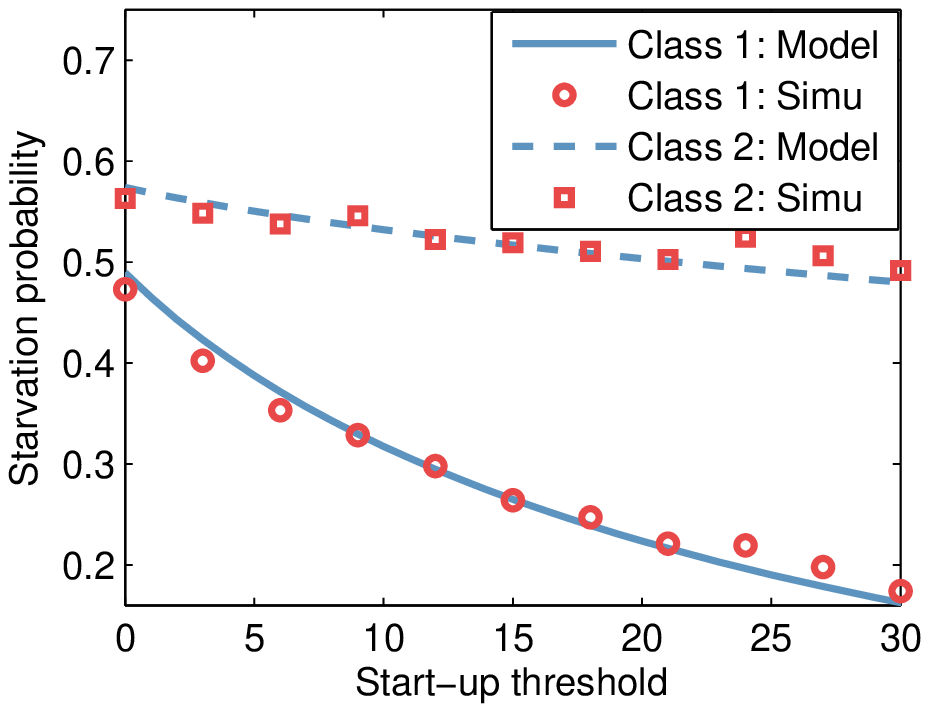}
\caption{Starvation probability  vs startup threshold with EPS discipline}
\label{fig:starvprob_threshold}
\end{minipage}
\begin{minipage}[t]{0.33\linewidth}
\centering
\includegraphics[width=2.5in]{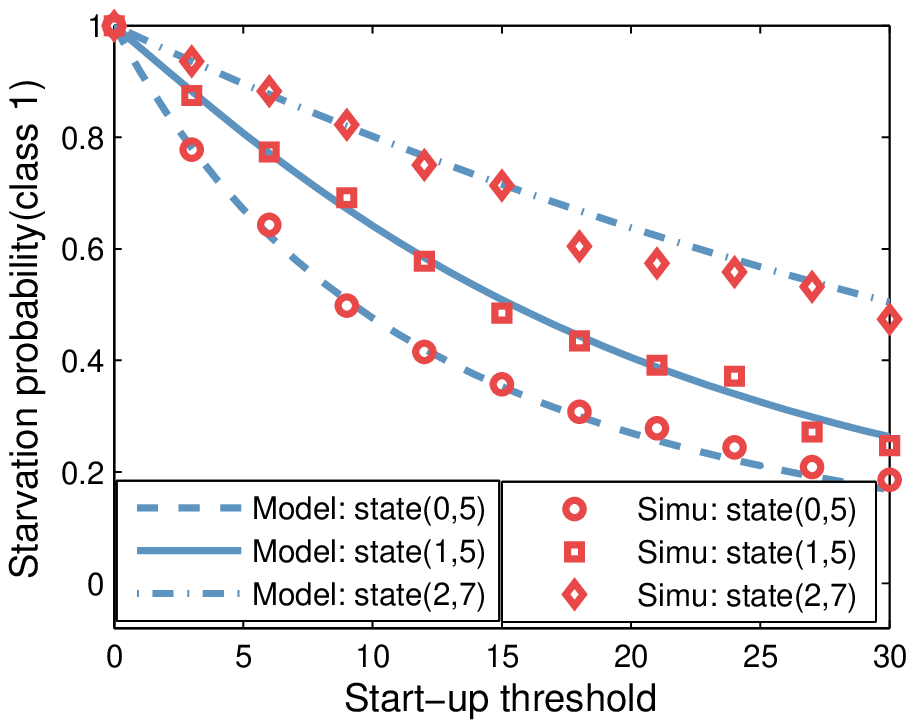}
\caption{Starvation probability (class-1)  VS startup threshold under different initial playback states.}
\vspace{-0.5cm}
\label{fig:starvation_states1}
\end{minipage}
\hspace{1ex}
\begin{minipage}[t]{0.33\linewidth}
\centering
\includegraphics[width=2.5in]{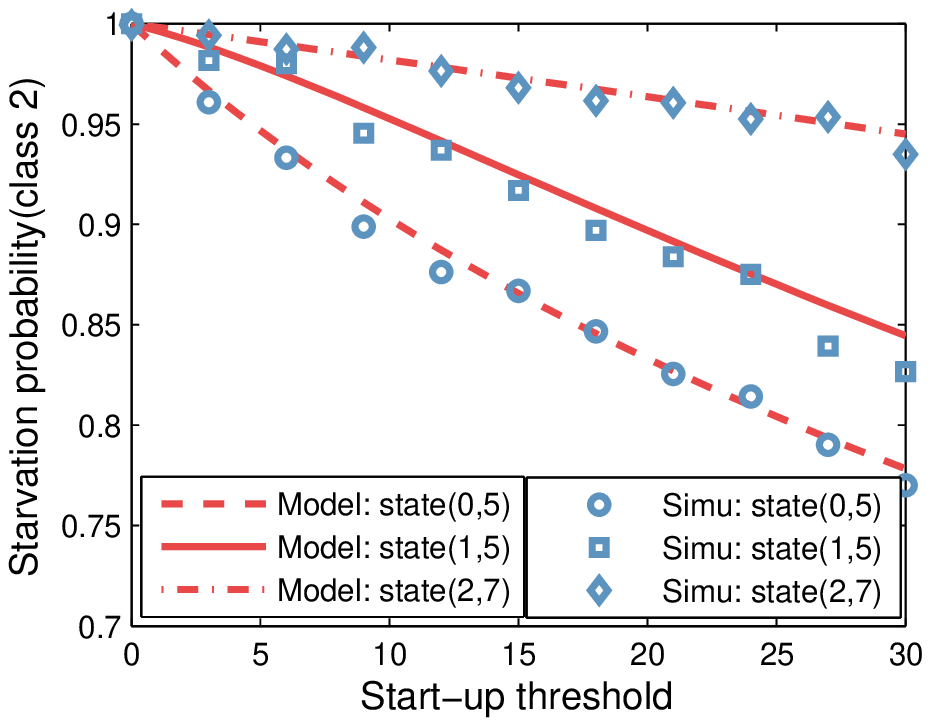}
\caption{Starvation probability (class-2)  VS startup threshold under different initial playback states.}
\label{fig:starvation_states2}
\end{minipage}
\vspace{-0.5cm}
\end{figure*}

Fig. \ref{fig:starvprob_threshold} illustrates the starvation probabilities of class-1 and class-2 flows as the start-up threshold increases from
0 to 30s. One can immediately observe that our mathematical model matches the trace-driven simulation well.
With the increase of start-up threshold, the starvation probability of class-1 flows decreases from 0.48 to 0.17, and that of
class-2 flows decreases from 0.57 to 0.48. Intuitively, the starvation probability of short views is more sensitive to the start-up
threshold than that of the long views.

We next evaluate starvation probabilities when the playback process begins at different states, as shown in Fig. \ref{fig:starvation_states1} and Fig. \ref{fig:starvation_states2}. State $(i,j)$ indicates when there are $i$ other class-1 flows and $j$ other class-2 flows. With more flows sharing the bottleneck, it is more likely to encounter the buffer starvation.

\begin{figure*}[!tb]
\begin{minipage}[t]{0.33\linewidth}
\centering
\includegraphics[width=2.5in]{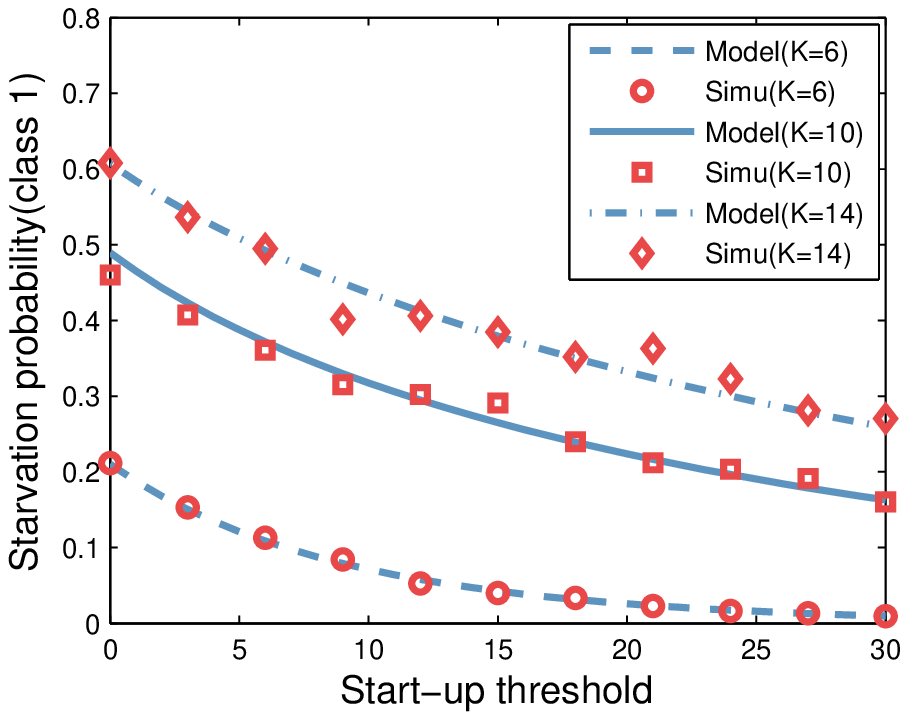}
\caption{Starvation probability (class-1)  VS startup threshold under different K.}
\label{fig:šstarvationK1}
\end{minipage}
\begin{minipage}[t]{0.33\linewidth}
\centering
\includegraphics[width=2.5in]{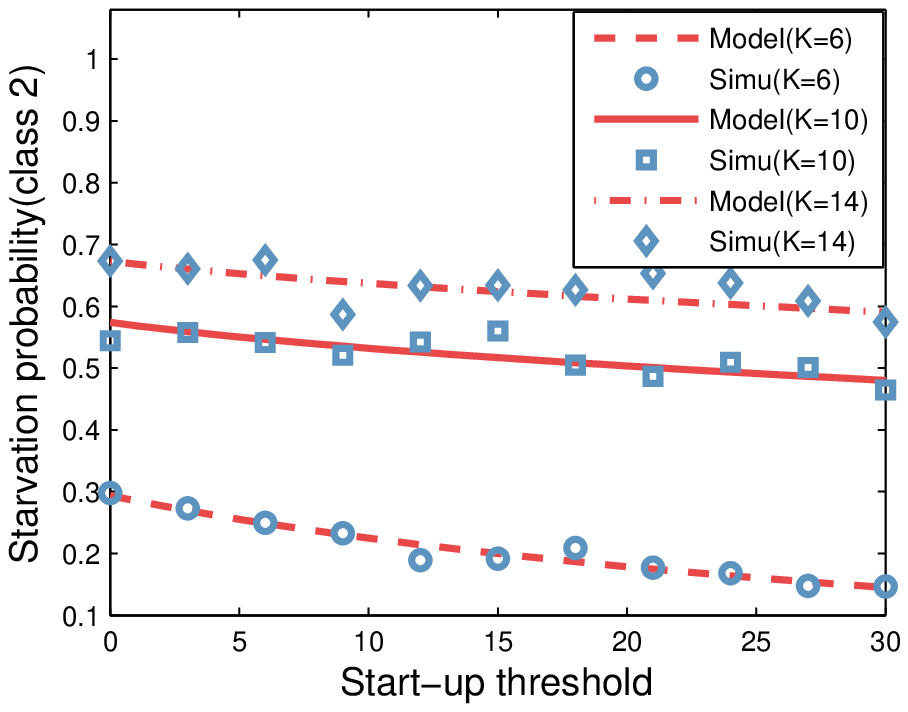}
\caption{Starvation probability (class-2)  VS startup threshold under different K.}
\label{fig:starvationK2}
\end{minipage}
\hspace{1ex}
\begin{minipage}[t]{0.33\linewidth}
\centering
\includegraphics[width=2.5in]{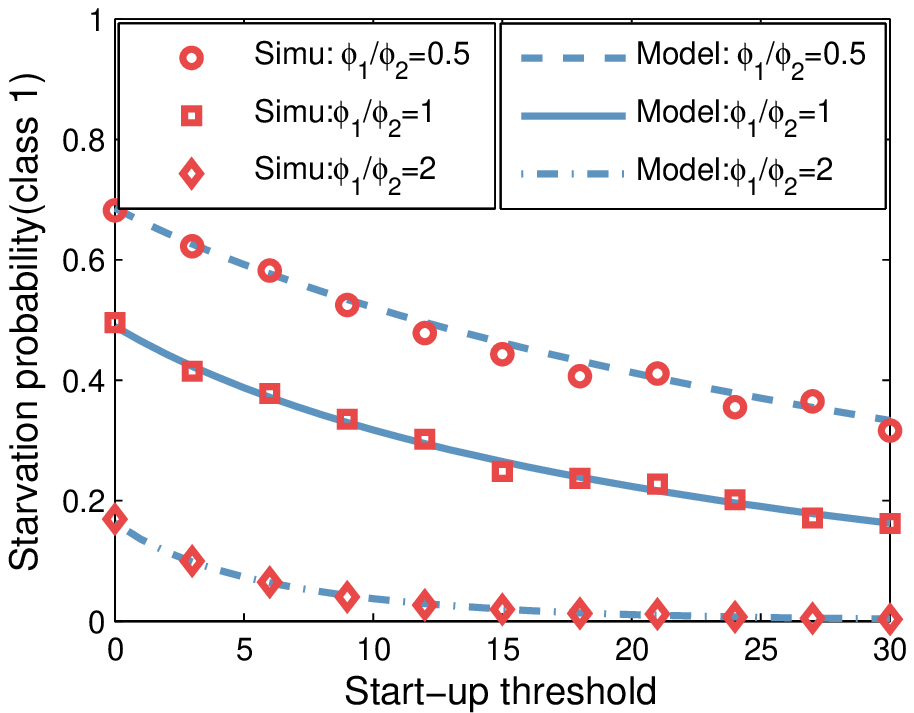}
\caption{Starvation probability  VS startup threshold with DPS discipline(class-1).}
\label{fig:starvprob_dps1}
\end{minipage}
\vspace{-0.5cm}
\end{figure*}

Furthermore, Fig. \ref{fig:šstarvationK1} and Fig. \ref{fig:starvationK2} compare the overall starvation probability of class-1 and class-2 flows with different maximum numbers of coexisting flows $K$, respectively. As $K$ increases, more flows (or congestions) may coexist at the BS, and hence resulting in higher starvation probabilities. Comparing two scenarios with $K=6$ and $K=10$,
both class-1 and class-2 flows exhibit strong dependence of starvation probability on the maximum number of flows.
Thus, our model could be used to design a proper admission control policy with tolerable starvation probabilities.

\begin{figure*}[!tb]
\begin{minipage}[t]{0.33\linewidth}
\centering
\includegraphics[width=2.5in]{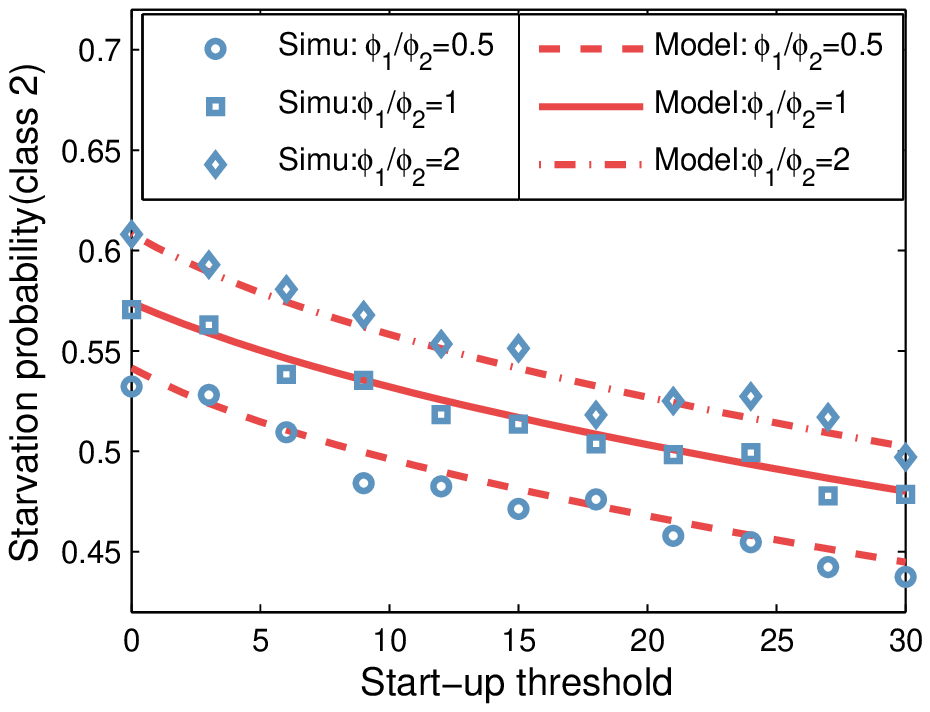}
\caption{Starvation probability  VS startup threshold with DPS discipline (class-2).}
\vspace{-0.5cm}
\label{fig:starvprob_dps2}
\end{minipage}
\hspace{1ex}
\begin{minipage}[t]{0.33\linewidth}
\centering
\includegraphics[width=2.5in]{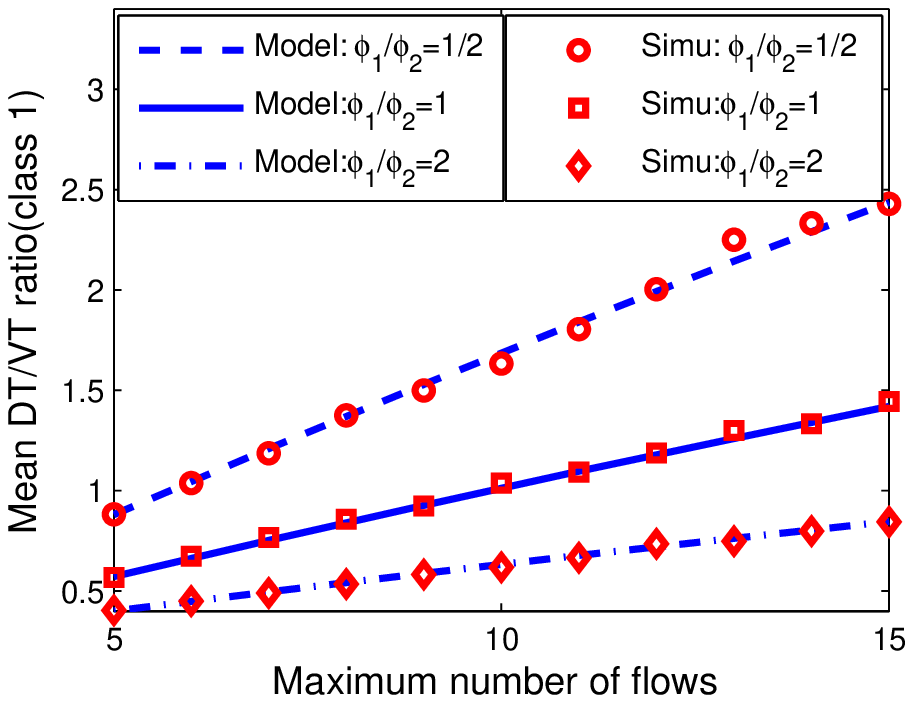}
\caption{Mean DT/VT ratio VS maximum number of simultaneous flows $K$ with DPS discipline (class-1).}
\label{fig:DT/VT_DPS1}
\end{minipage}
\hspace{1ex}
\begin{minipage}[t]{0.33\linewidth}
\centering
\includegraphics[width=2.5in]{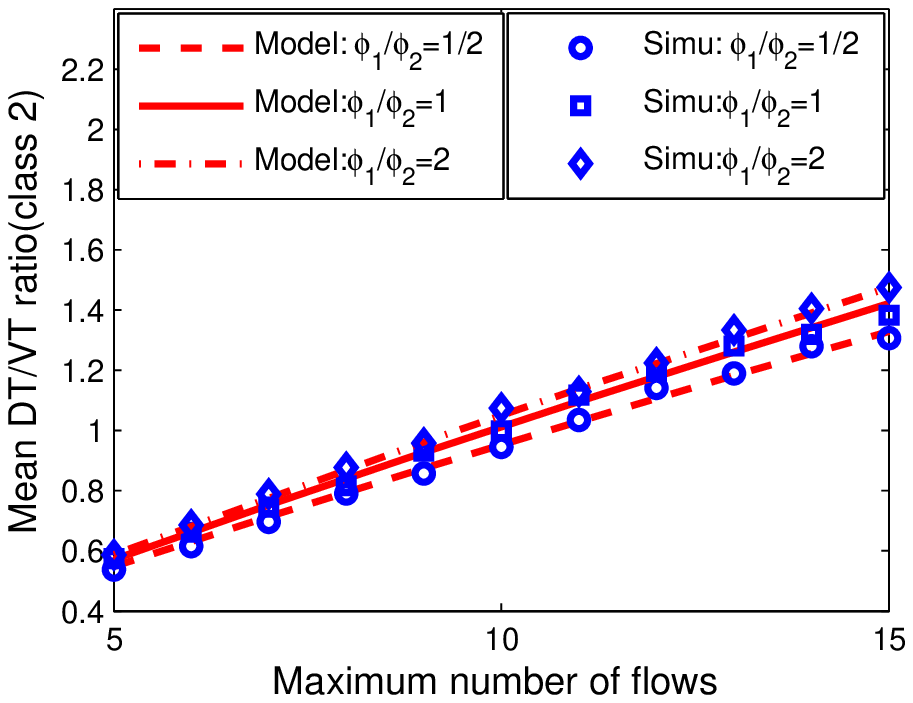}
\caption{Mean DT/VT ratio VS maximum number of simultaneous flows $K$ with DPS discipline (class-2).}
\vspace{-0.5cm}
\label{fig:DT/VT_DPS2}
\end{minipage}
\vspace{-0.3cm}
\end{figure*}

We next evaluate the impact of scheduling priority on the starvation behaviors. As shown above, the short and long views
exhibit different starvation behaviors. Our purpose here is to investigate the effect of modifying scheduling priorities on
different class of flows. Fig. \ref{fig:starvprob_dps1} and Fig. \ref{fig:starvprob_dps2} plot the starvation probability of class-1 and class-2 flows when the ratio of
scheduling weights $\frac{\phi_1}{\phi_2}$ is 2, 1 and $\frac{1}{2}$ respectively. Let us take the case $\frac{\phi_1}{\phi_2} = 1$ as
the benchmark. When $\frac{\phi_1}{\phi_2}$ is 2, the starvation probability of class-1 flows
decreases significantly at the cost of a slight increase in the starvation probability of class-2 flows.
On the contrary, when $\frac{\phi_1}{\phi_2}$ is $\frac{1}{2}$, the starvation probability of class-1 flows increases by nearly 20\%, while
the starvation probability of class-2 flows has a marginal reduction less than 5\%.

\textbf{Mean DT/VT ratio.} The DT/DV ratio is an important metric that reflects the best buffering ratio during the entire video playback.
In Fig.\ref{fig:DT/VT_DPS1} and \ref{fig:DT/VT_DPS2}, we plot the mean DT/DV ratio as the
admission control threshold $K$ increases from 5 to 15.
The increase of $K$ leads to a higher mean DT/DV ratio, which implies that a lot of time is spent on the video buffering.
An interesting observation is that under EPS scheduling (i.e. $\frac{\phi_1}{\phi_2}=1$),
the mean DT/VT ratios of the short views and long views are the same.
This is because that the per-flow throughput of all the on-going flows are identical no matter that they are short or long views.
In the heavy traffic regime, a long view will encounter a larger buffering time than a short view. However, the relative buffering ratio may
be very close.

In Fig.\ref{fig:DT/VT_DPS1}, we further compare the mean DT/VT ratios of class-1 views
under different DPS scheduling schemes. When the scheduling weights have $\frac{\phi_1}{\phi_2}=2$, the
mean DT/VT ratio of class-1 views reduces by nearly 50\%. When a higher priority is offered to class-2 views,
the mean DT/VT ratio of class-1 views increases nearly 100\%.
Fig.\ref{fig:DT/VT_DPS2} illustrates the impact of scheduling weights at the BS on the mean DT/VT ratio of class-2 views.
Similarly, we evaluate three scenarios, i.e. $\frac{\phi_1}{\phi_2} = \frac{1}{2}, 1$ and $2$. For class-2 views,
their mean DT/VT ratio only changes by less than 10\%. Offering a higher priority to class-2 flows results in a
slight improvement in the mean DT/VT ratio.

\noindent\textbf{Remark:} Our modeling framework provides a fundamental understanding on the QoE tradeoff of
heterogeneous video streams. When short views and long views possess different user perceptions toward buffer
starvations (e.g. starvation probability or DT/VT ratio), the BS
is able to configure different scheduling priorities to each class of views so that the overall QoE can be improved.
For instance, a higher scheduling priority can be given to short views if they are more sensitive to the probability
of buffer starvation.

\section{Refining QoE Models with Realistic Considerations}
\label{sec:extension}
New challenges are introduced in the modeling of QoE metrics when progressive downloading is adopted in realistic Internet streaming services. In this section, we generalize our analytical framework with the consideration of progressive downloading.

\subsection{The Role of Progressive Downloading}

In our basic QoE models, the viewing time is taken to estimate the traffic load, and the accuracy of mathematical models are
proven by trace-driven simulations. However, the media player does not know how long a user will watch in advance, thus
usually downloading more data than been watched. This implies that using viewing time tends to underestimate the traffic load.
Today's video streaming service is mostly pull-based, which means that the user requests video chunks, and the server responds
passively. To ensure smooth video playback, the user continuously requests a new video chunk as soon as the previous one
has been downloaded, which it is called ``progressively downloading (PD)''. In comparison to the controlled bandwidth provision,
PD is able to reduce the risk of playback interruption. A common criticism to PD is
that it leads to considerable waste traffic because a fraction of video content is downloaded but not consumed.
We revisit these arguments in a different scenario where multiple video streams share the same bottleneck.
Intuitively, when PD strategy is implemented, some users may download more video content than they watch.
Due to the limited bandwidth, some other users may incur smaller downloading speeds so that they are prone to encounter starvations.
We can speculate that PD will exacerbate the starvation behaviors.
Whereas it is unclear to what extent the QoE metrics are influenced, and which class of flows is more heavily influenced.
A natural question is whether our analytical framework can be modified to incorporate PD in practice.

\begin{figure*}[!tb]
\begin{minipage}[t]{0.33\linewidth}
\centering
\includegraphics[width=2.5in]{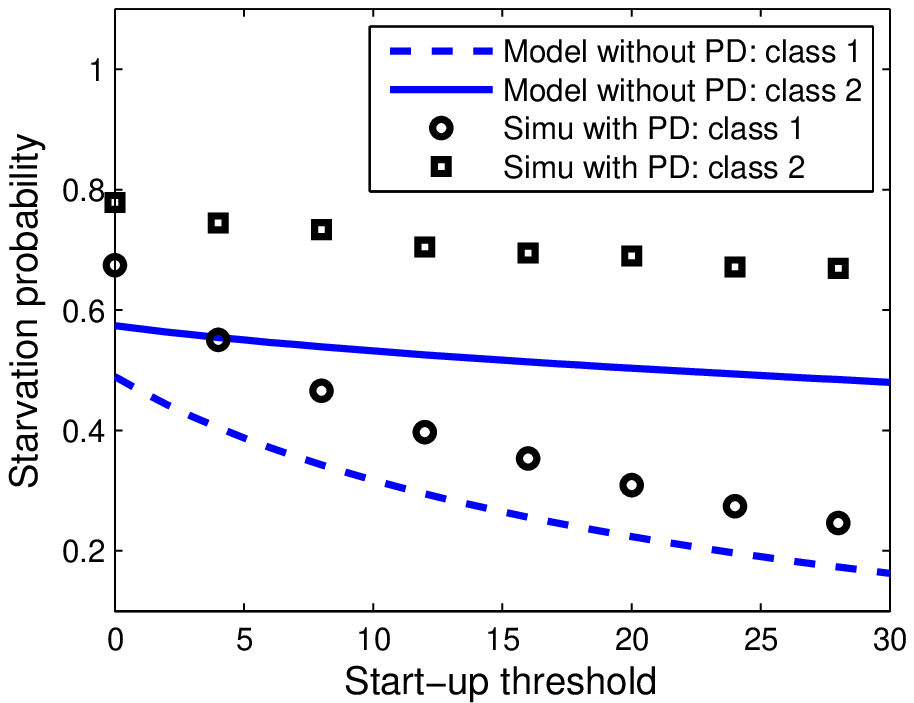}
\caption{The role of progressive downloading with traffic load {$\rho=0.96$}.}
\label{fig: role_of_PD2}
\end{minipage}
\hspace{1ex}
\begin{minipage}[t]{0.33\linewidth}
\centering
\includegraphics[width=2.5in]{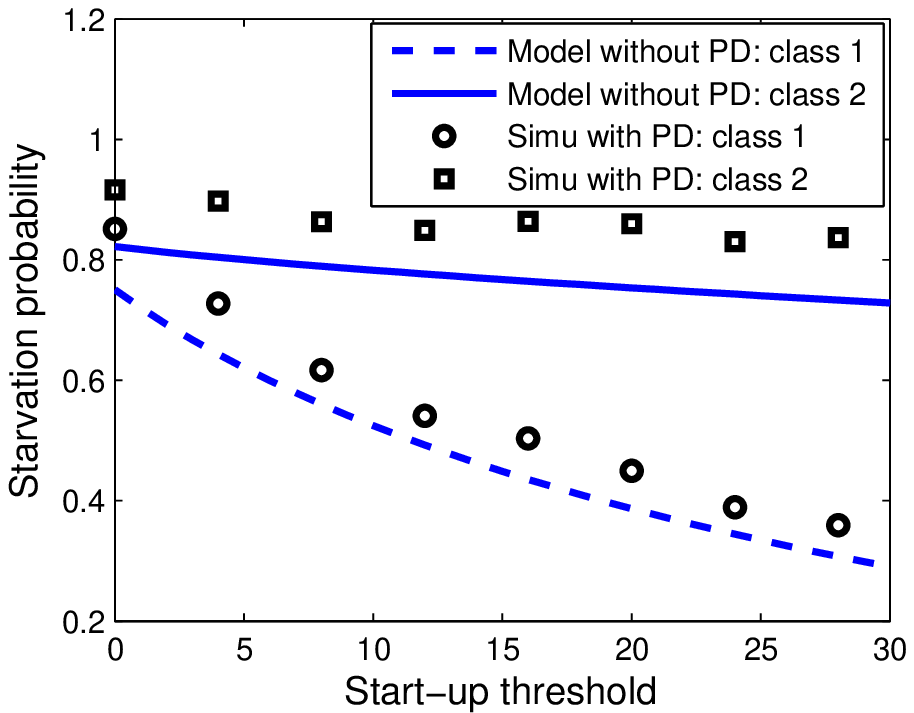}
\caption{The role of progressive downloading with traffic load {$\rho=1.2$}.}
\label{fig: role_of_PD3}
\end{minipage}
\hspace{1ex}
\begin{minipage}[t]{0.33\linewidth}
\centering
\includegraphics[width=2.5in]{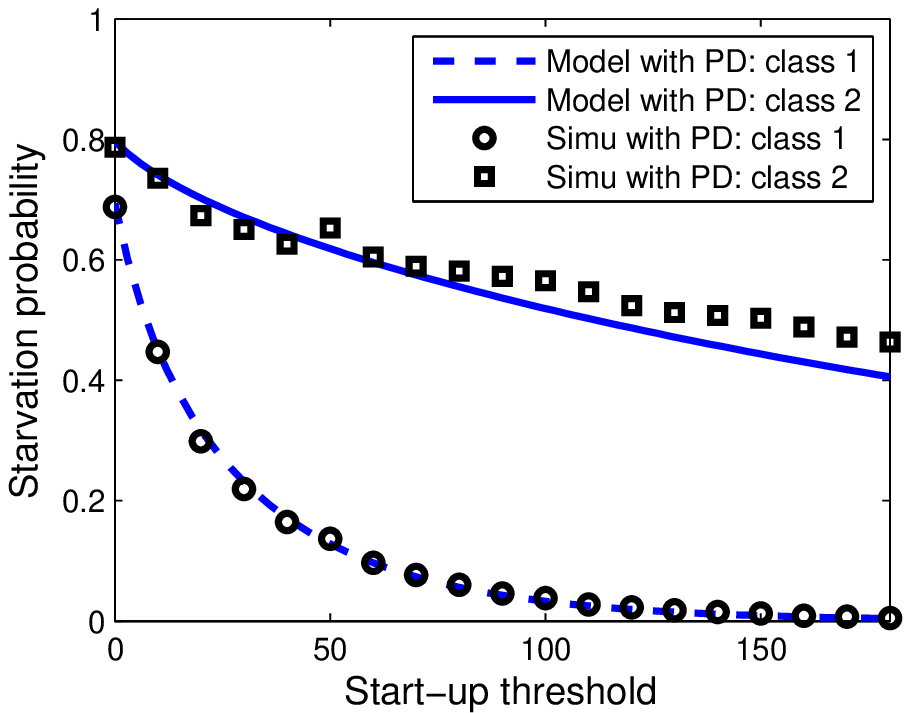}
\caption{Starvation probability with the PD scheme and without consideration of video duration: model vs simulation}
\label{fig:starvprob_pd_novlength}
\end{minipage}
\vspace{-0.3cm}
\end{figure*}

We first conduct more experiments to evaluate the influence of progressive downloading on the starvation probabilities.
Our trace-driven experiments are configured the same as those in Section \ref{sec:simu} except that
PD is implemented.
Fig.\ref{fig: role_of_PD2} shows that the starvation probability increases apparently above the basic model,
which deteriorates the accuracy of the basic model.
Intuitively, the root-cause of his mismatch is that the viewing time cannot accurately reflect the traffic load.
When the traffic load is inside the capacity region (i.e. $\rho<1$), the downloading rate will surpass the video bitrate at certain states.
Hence, some flows (especially with long video durations) are likely to buffer more waste data than that will be consumed.
Due to the invariable total bandwidth,
some other flows have to download less data than that will be watched. Later on, when the system state transits to a more congested one,
these flows may encounter buffer starvation more easily.
Therefore, the basic model only provides a loose lower bound of the QoE performance, as shown in Fig. \ref{fig: role_of_PD2}.
We next evaluate the impact of progressive downloading on the starvation probability in Fig.\ref{fig: role_of_PD3}
when the traffic load overrides the capacity (e.g. $\rho=1.2$ at the basic model).
In this circumstance, the network is very congested so that there are less chances for the flows to download more waste data.
Then, the basic model can provide an accurate estimation of the QoE metrics.

Besides, the influence of PD on the QoE is mainly brought be the abandonment of long videos with short viewing times.
The short videos are more likely to be finished and there will be very little wasted data.
With PD, a user of a long video may download more data when the bottleneck is shared by a small number of
concurrent users. While the early departure of this user will cause non-negligible waste of bandwidth.

The modeling of QoE metrics is very \textbf{challenging} when the PD scheme is considered.
The fundamental reason is that the volume of fetched video content cannot be quantified, even the distribution of viewing time is known
as a priori. There are three root causes.
\begin{itemize}

\item Firstly, the amount of unwatched content depends on the throughput of this flow (or depends on the number of coexisting
flows equivalently). For example, the PD scheme downloads more video content that is not watched
at state $(0, 0)$, and less at state $(0, K{-}1)$.

\item Secondly, the PD scheme is coupled with not only the
viewing time but also the video duration.
Let us take two representative videos A and B as examples. Video A has a duration of 1000 seconds and has been watched by a user for
200 seconds. Video B has a duration of 50 seconds, and has been watched for 40 seconds.
The user with PD will download at most 10 seconds when watching video B, while can download
much more unwatched video content.

\item Thirdly, the service time distribution is correlated with the start-up threshold.
For instance, we suppose that the content provider sets the start-up threshold to be 20 seconds
for a video with the duration of 200 seconds. A user watches 10 seconds video content. Then,
he downloads 10 seconds video content without the PD scheme, and downloads at least 20 seconds video content
with the PD scheme. The start-up threshold is equivalent to a dead zone added to the viewing time of each class of
video content. Even though the viewing times are exponentially distributed in each class, the service times do not follow
the exponentially distributions any more.
\end{itemize}
As an outcome, precise models of QoE metrics are intractable in the presence of the PD scheme.
Among the above three factors, the first one imposes the main difference between PD and none-PD scenarios.
The remaining two factors exert additional constraints with regard to the video duration.
We extend our analytical model to study the QoE metrics by taking account of the first factor, and later on we will interpret
the reason why the other two factors have less prominent impacts.

\subsection{Extension to Progressive Downloading Scenario}

Due to the affect of progressive downloading, we further propose a refined model. The PD scheme basically changes the departure rates.
In the basic models, the viewing time is treated as the amount of video content that will be downloaded.
knowing the viewing time, the tagged user leaves the BS immediately after downloading the video content that will be watched exactly.
When the PD scheme is considered, we examine the dynamics of playout buffer of the tagged flow in two different conditions.

\noindent i) When the throughput of the tagged flow is \textit{less than} the video playback rate at state $(i, j)$,
the buffer length will decrease.
The PD scheme does not lead to more unwatched video content. Given the exponentially
distributed viewing time for each class of flows, the transition rate from state $(i, j)$ to state $(i{-}1, j)$ (or from $(i, j)$ to $(i, j{-}1)$) is
the same in both the PD and the none-PD schemes. Similar argument also holds in the transition rate from
state $(i, j)$ to the absorbing state.

\noindent ii) When the throughput of the tagged flow is \textit{greater than} the video playback rate at state $(i, j)$,
the buffer length increases.
However, the tagged user has a tendency to download more content than watched so that the transition rate from
state $(i, j)$ to state $(i{-}1, j)$ becomes smaller.
At state $(i, j)$, the mean service time of the class-1 tagged flow is not $\frac{C\phi_1 \theta_1}{Bitrate (\phi_1 i + \phi_2j )}$, but is exactly the viewing time $1/\theta_1$. Therefore, the transition rate from state $(i, j)$ to state $(i{-}1, j)$ is $i\theta_1$. Similarly, the transition rate
from state $(i, j)$ to the absorbing state is computed as $\theta_1$.

Formally, we derive the new state transition rates as follows:
\begin{eqnarray}
\mu{'}_{i,j}&=&\min\{\mu_{i,j}, i\theta_1\}, \nonumber  \\
\nu{'}_{i,j}&=&\min\{\nu_{i,j}, j\theta_2\} , \nonumber \\
\varphi{'}_{i,j}&=&\mu{'}_{i,j}/i.  \nonumber
\end{eqnarray}
In this new scenario, all the QoE models remain unchanged
except for the transition rates $\mu_{i,j}$, $\nu_{i,j}$ and $\varphi_{i,j}$ substituted by $\mu{'}_{i,j}$, $\nu{'}_{i,j}$ and
$\varphi{'}_{i,j}$ for all $(i, j)\in S$.

To validate the effectiveness of our analytical framework, we present a set of trace-driven simulations. Note that the video duration
is not considered. We only extract the real viewing times from the dataset while assuming the video durations to be long enough.
Our purpose is to isolate the impact of video duration on the modified QoE models.
To avoid any inconsistence, we continue to adopt the same wireless setting with traffic load $\rho =0.96$ and the
scheduling weights $\frac{\phi_1}{\phi_2}=1$.
Fig. \ref{fig:starvprob_pd_novlength} compares the refined model with the simulation result of starvation probability
when the PD scheme is considered. For the class-1 flows, the refined model matches the simulation result very well, and the
accuracy is almost not influenced by the start-up threshold. For the class-2 flows, the refined model is accurate when
the start-up threshold is less than 80 seconds. As the start-up threshold further increases, the gap between the refined model and
the trace-driven simulation increases accordingly. When the start-up threshold is 180 seconds, this gap grows to nearly 8\%.
We interpret our observations as follows. When the PD scheme is considered, the existence of start-up threshold breaches
the exponential distribution of viewing times in each class of flows.

The service time of a flow is actually the start-up threshold plus the viewing time,
while the refined model only uses the viewing time as the service time.
Let us suppose that the startup threshold and the viewing time of class-1 flow are denoted by $t_{th}$ and $t_{vt}$ respectively.
The realistic service time is $t_{th}+t_{vt}$. This explains the gap between the refined models and simulation results.
Lack of the consideration of start-up delay, the traffic load is underestimated in the refined model.
Therefore, the starvation probability computed by our refined model is lower than the simulation result.
The gap is negligible when the start-up threshold is small, and enlarges as the start-up threshold increases.
This phenomenon can be explained by noting that the difference between
$t_{th}+t_{vt}$ and $ t_{vt}$.
However, it is extremely difficult to further integrate the coupling effect of start-up threshold and PD scheme in our refined model.
Here, $t_{vt}$ is exponentially distributed while $t_{th}$ is a constant so that $t_{th}+t_{vt}$ no longer follows the exponential distribution. Taking the start-up threshold into account, the dynamics of the number of flows at the BS is no Markovian so that our analytical
framework does not apply. The optimistic observation from Fig. \ref{fig:starvprob_pd_novlength} shows that the
gap is very small when the startup threshold is below 100 seconds. Therefore, it is not necessary to
further sacrifice the complexity of the model for precision.

\subsection{QoE Model with Consideration of Video Duration}

As mentioned earlier, both the refined models and trace-driven simulations in Fig.\ref{fig:starvprob_pd_novlength} assume that
the video durations are long enough. In a realistic system, progressive downloading can not proceed without a boundary.
Even when the network condition is ideal, the media player can at most buffer the whole video.

It is extremely difficult to integrate the variable video duration into our refined model.
For each video, there are two correlated random variables, the viewing time and the video duration.
It is unclear whether a flow leaves the
BS because of the early departure or the completion of downloading the entire video.
Though our refined model cannot perfectly combine all the factors, we experimentally show that it is a good estimation of
realistic streaming QoE, and explain the reason.

\begin{figure*}[!tb]
\begin{minipage}[t]{0.33\linewidth}
\centering
\includegraphics[width=2.5in]{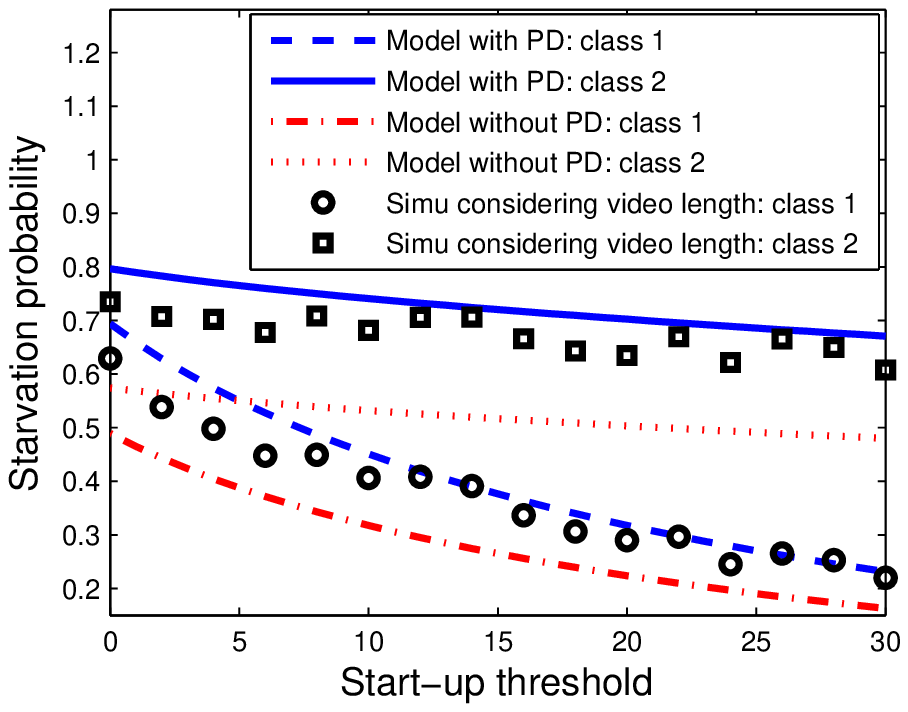}
\caption{Simulation results with consideration of video length and PD, compared to the basic models and refined models(traffic load {$\rho=0.96$}).}
\label{fig:starvprob_videoduration1}
\end{minipage}
\hspace{1ex}
\begin{minipage}[t]{0.33\linewidth}
\centering
\includegraphics[width=2.5in]{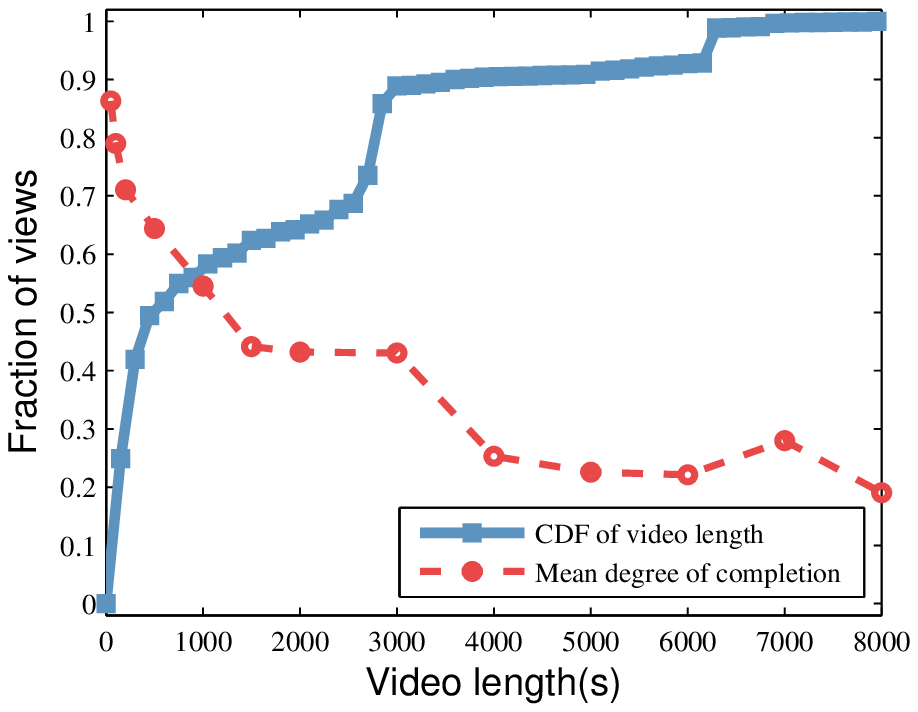}
\caption{The CDF of video length and the mean degree of completion.}
\vspace{-0.5cm}
\label{fig:cdf_videolength}
\end{minipage}
\hspace{1ex}
\begin{minipage}[t]{0.33\linewidth}
\centering
\includegraphics[width=2.5in]{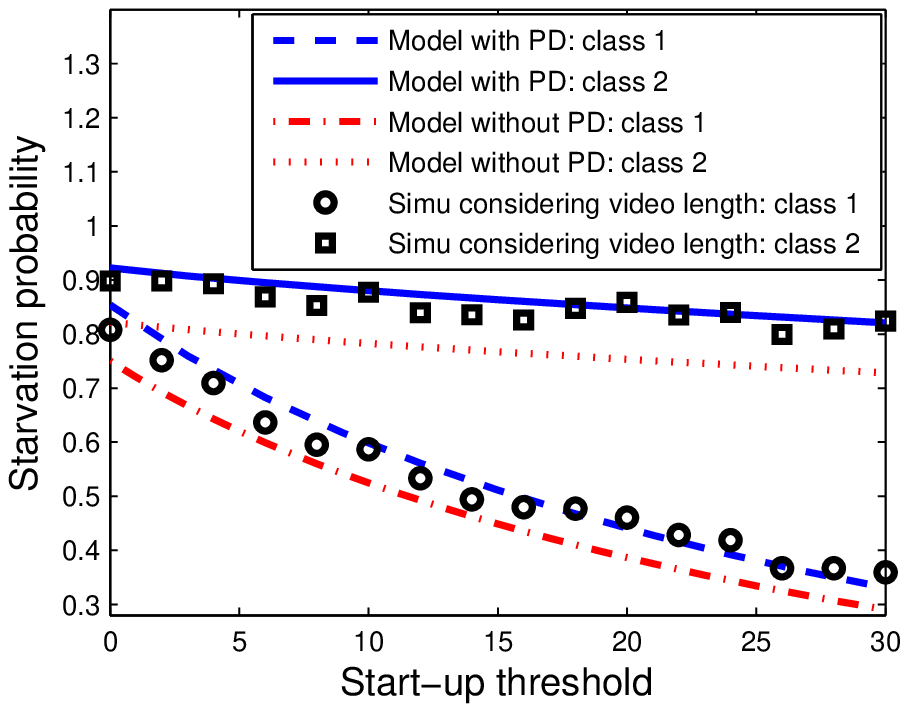}
\caption{Simulation results with consideration of video length and PD, compared to the basic models and refined models(traffic load {$\rho=1.2$}).}
\label{fig:starvprob_videoduration2}
\end{minipage}
\vspace{-0.3cm}
\end{figure*}

We conduct a set of simulations to show the starvation probabilities with the consideration of video duration.
In Fig.\ref{fig:starvprob_videoduration1}, we compare the refined model with trace-driven simulation on the starvation probability
when the scheduling weights are identical. The remaining parameters including the flow arrival rates, the viewing times and the
bitrate are the same as those in the preceding experiments.
The refined model seemingly overestimates the starvation probability in almost all the experiments.
According to the refined model, the playout buffer of a user continues to increase if the throughput is greater than the bitrate.
In practice, the playout buffer stops to grow if the video has been downloaded completely. Then, the refined model
overestimates the amount of video content fetched by the PD scheme. Therefore, the starvation probability computed by the
refined model can be deemed as an upper bound of realistic starvation probability.
Furthermore, we find that the refined model can well predict the
starvation probabilities obtained from experiments. Their gap ranges from 0 to 7\%, and is around 7\% for class-1 views and around
5\% for class-2 views. To cast about for the reason of this inaccuracy, we need to examine the difference between the viewing
time and duration for each video. Fig.\ref{fig:cdf_videolength} plots the cumulative distribution function of the video duration from our dataset.
The average completion ratio, defined as the quotient of the viewing time over the video duration, is also illustrated. 
For long videos, their completion ratios are usually small so that the assumption of infinite video duration holds in the refined model.
However, for short videos, their completion ratios are usually large so that the PD scheme will not be able to prefetch a lot of
video content even under the less congested scenarios. This does not conform to the assumption of the refined model, thus
causing to the gap between the model and the simulation. We continue to evaluate the accuracy of the refined model with a traffic load
$\rho = 1.2$, i.e. a more congested scenario, in Fig.\ref{fig:starvprob_videoduration2}.
One can see that the influence of PD is weaken when the network is more congested, and our refined model serves as
a good approximation of QoE with the consideration of mixed practical features including PD, start-up threshold and finite video duration.

\section{Predicting Viewing Time with "Least" Information}
\label{sec:prediction}

In this section, we address the practical challenge that network operators are unaware of the viewing time so to hinder any QoE
enhancement strategy.

\subsection{Priori Mapping Between Viewing Time and Video Duration}

Although our study presents important insights on the QoE modeling and optimization, the network operators cannot directly
use our analytical framework to design new resource allocation strategies. The main challenge is that they do not acquire
the necessary information to perform the computation. With appropriate packet inspection and flow inspection approaches, the
network operator can accurately measure the flow arrival rate of each class. However, the proposed framework relies on the differentiation
of classified videos (i.e. short and long viewing times), which is unknown at the BS.
Estimating the viewing time encounters two practical difficulties. On one hand, estimation of viewing time may
require a lot of attributes such as video type, video content, and individual information of a viewer, etc.
These information is proprietary to the content providers who are unlikely to release them for the sake of privacy and commercial
concerns. On the other hand, the viewing time of a video request is unknown before this video has been played.
Hence, the network operators need to deliberate what is the ``least'' information they can obtain from the content providers
so as to predict streaming QoE. It is the necessary condition to put our analytical framework into practice.

We hereby propose a simple but effective approach to predict
the viewing time with almost the least information, in which only the video duration and the a small set of CDFs of completion ratio
are needed. According to Fig.\ref{fig:cdf_videolength}, when the video duration increases, the completion ratio decreases monotonically.
Therefore, we conjecture that the video duration can be used to infer the viewing time. Given the estimation of viewing time, we
can further speculate the class of the incoming video request. To this goal, the first step is to build an appropriate mapping
between the viewing time and the video duration.

Let $t_{view}$ and $t_{video}$ denote the viewing time and video length. We define \emph{Degree of Completion (DoC) $\omega$}  as $\frac{t_{view}}{t_{video}}$, while $\omega$ equaling 1 indicates the video is watched completely. With the knowledge of requested video's length, $\omega$ can reflect how long the video will be watched.  It is an intuitive thought that the longer one video is, more likely it will be that the user may quit watching it halfway. This suggests that there should be some relationship between the video length and the \emph{DoC}.

Therefore, we divide videos into several classes according to their length. We study the distribution of $\omega$ in each class . By calculating $\omega$ for all views, we can plot CDFs of $\omega$ in each class, as Fig.\ref{fig:doc} shows. We can see that videos with different video length have different CDF of $\omega$ and the longer the videos are, the lower the degree of completion is likely to be, which inspires us to capture the degree of video completion distribution by curve fitting. From Fig.\ref{fig:doc}, we can see that for $\omega$ ranging from to 0 to 95\% (taking video length 0-100s for example), the CDF follows the Power Law distribution. Thus, we can capture the distribution of these $\omega$ by fitting a  curve
\begin{equation}
F(\omega)=c\omega^{-a}.
\end{equation}

Besides this region of $\omega$, there remains a short span of $\omega$, where the CDF can be considered to be linear. So we assume $\omega$ is uniformly distributed within this short span, denoted by $\epsilon$. Then, the overall degree of completion distribution can be fitted by
  \begin{equation}
F(\omega)=\left\{
\begin{array}{rcl}
       c\omega^{-a} \quad \quad \quad &      & {0 < \omega \le 1.05-\epsilon}\\
\frac{1-c(1.05-\epsilon)^{-a}}{\epsilon} \omega-b &      & {1.05-\epsilon < \omega \le 1.05}
\end{array} \right.
 \end{equation}
Using similar method in Section \ref{sec:measurement}, the parameters can be obtained, as shown in Table
\ref{table:mapping_viewtime_duration} (taking the interval 0-50 seconds as an example). We divide the video duration into nine intervals empirically that will strike a balance between
the accuracy of estimation and the complexity.
\begin{table}[!tbp]
\centering
\begin{tabular}{l|c|c|c}
\hline
$\epsilon $& c & a & b\\ \hline
 0.9460 & 0.2733 & 0.6383 & 6.723 \\  \hline
\end{tabular}
\caption{Parameters of fitting curves for completion ratio model (0-50 seconds).}
\label{table:mapping_viewtime_duration}
\vspace{-0.5cm}
\end{table}

Although the content providers are reluctant to release the meta information on videos and private information on streaming
users, they are likely to inform the distribution of completion ratio offline and the video duration online to the network operators.
Among the meta information concerning a video, the video duration is almost of the least value, and is often publicly available.
The content providers only need to offer a set of CDF curves in Fig.\ref{fig:doc}, which is regarded as the ``least'' information.
We believe that our method provides a feasible way for the collaboration of the network operators and content providers to
jointly optimize QoE of video streaming.

\subsection{Bayesian Inference of Viewing Time}

The nine CDF curves of completion ratio are obtained from a large-scale dataset. Actually, they are a set of compressions
of the original viewing behavior. Meanwhile, the CDF curves may vary on different days in a week.
An interesting question is whether a very small number of CDFs can be employed to
infer the viewing time reversely. The estimation of viewing time is based on the priori knowledge of the distribution of completion ratio,
which we name this approach ``Bayesian inference''. When a new video stream is requested, the content provider can notify
the video duration to the network operator. With this knowledge, the network operator adopts a simple table look-up to find
the distribution of DoC and generates a random DoC value. According to this DoC value, he will
randomly generates a viewing time. Once the viewing time has been estimated, the network operator can 
further use Eq.\eqref{eq:bayesian} to compute the probabilities of each view belonging to class-1 and class-2 respectively. 
%
%
%

\begin{figure*}[!tb]
\begin{minipage}[t]{0.33\linewidth}
\centering
\includegraphics[width=2.5in]{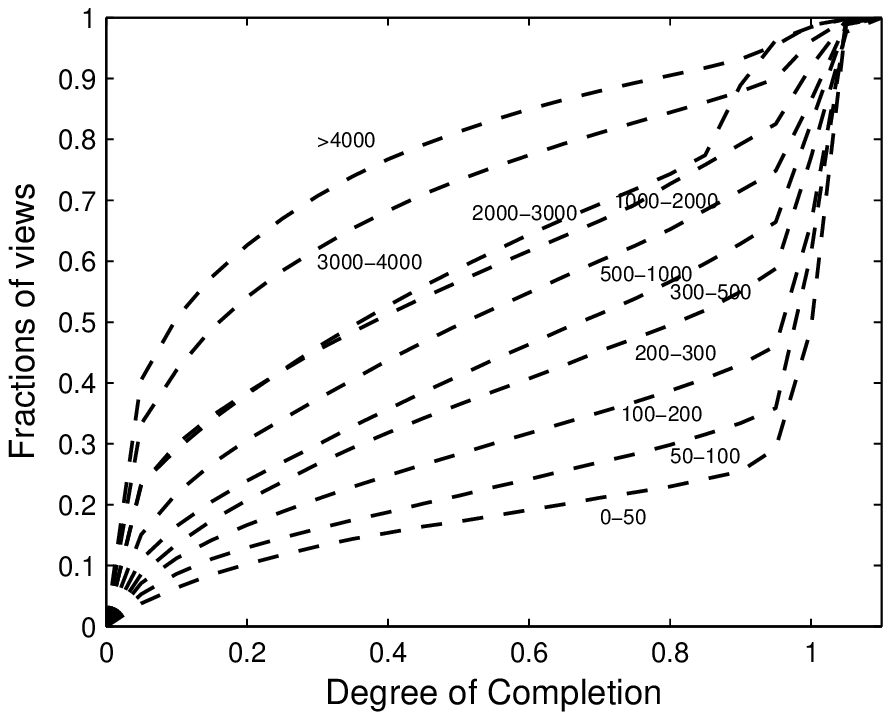}
\caption{CDF of videos' DoC (Degree of Completion).}
\label{fig:doc}
\end{minipage}
\hspace{1ex}
\begin{minipage}[t]{0.33\linewidth}
\centering
\includegraphics[width=2.5in]{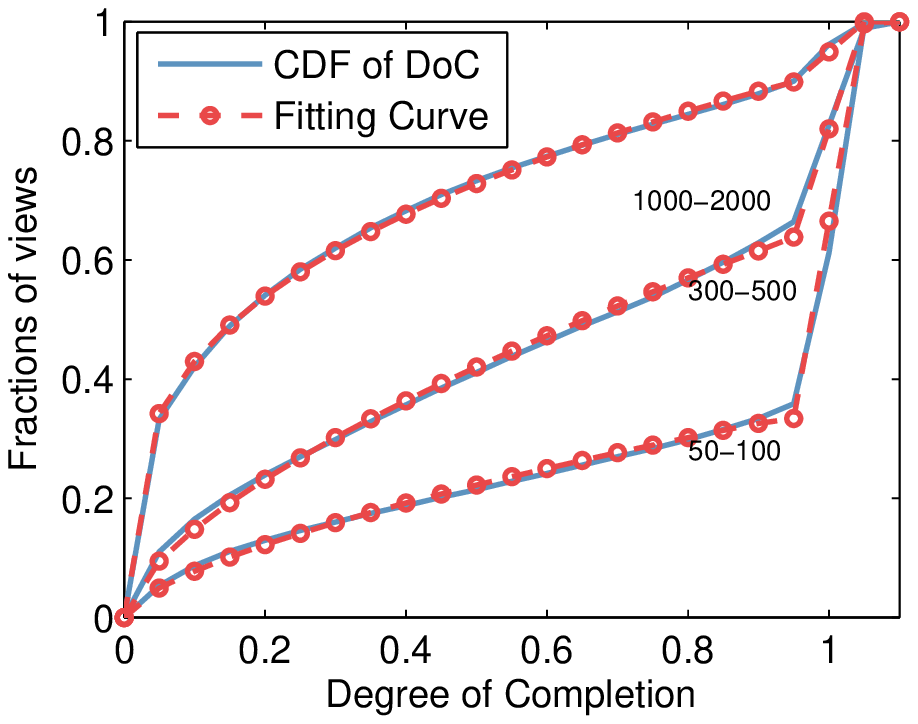}
\caption{Bayesian inference of viewing time distribution.}
\label{fig:bayesian1}
\end{minipage}
\hspace{1ex}
\begin{minipage}[t]{0.33\linewidth}
\centering
\includegraphics[width=2.5in]{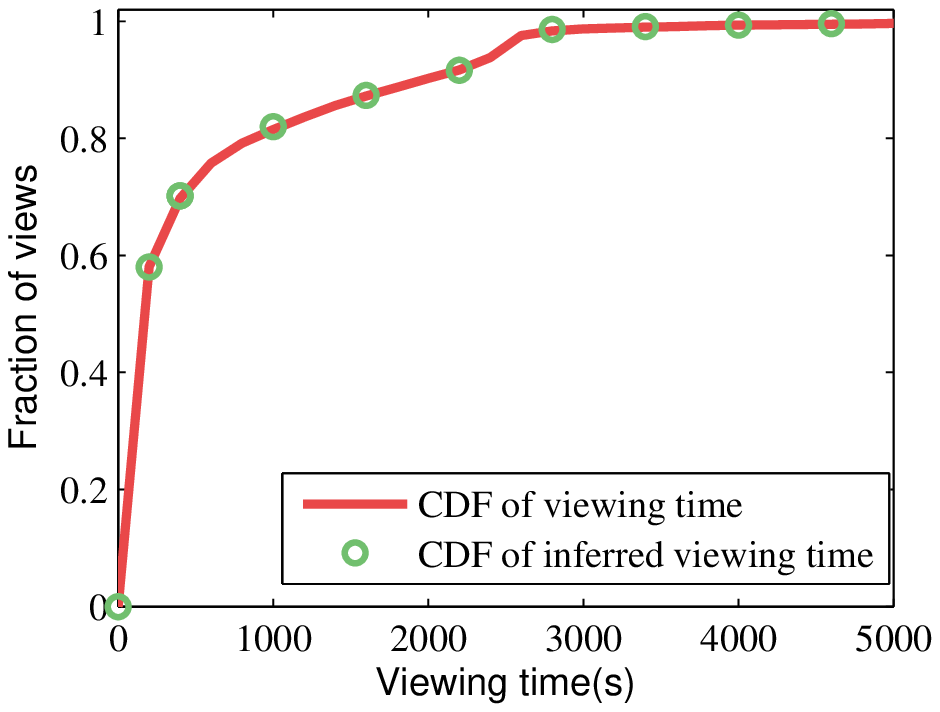}
\caption{Reconstructing viewing time distribution compared with original distribution in the dataset}
\label{fig:bayesian2}
\end{minipage}
\vspace{-0.3cm}
\end{figure*}

A subsequent question is whether our Bayesian inference approach can accurately reflect the distribution of viewing time.
Fig.\ref{fig:bayesian1} takes three video length intervals, 50-100 seconds, 300-500 seconds and 1000-2000 seconds,
to show the accurate fitting results.
Similarly, the fitting curves for videos in other classes can also be obtained. These distributions of DoC are statistical results based on the video views in one day. After verification, we find that these distribution don't change over time, which means curve fitting is a one-time effort and the parameters can be stored by BS to infer the DoC of following video.  When new video request arrives, it can be categorized according to its length. Then BS can generate the DoC for this video  randomly based on its corresponding CDF. Thus, viewing time $t_{vt}$ can be obtained by multiplying DoC by video length $t_{video}$. Fig.\ref{fig:bayesian2} shows that statistically, viewing time can be inferred accurately in this way. The CDF of inferred viewing time is almost the same as the CDF of true viewing time.

\section{Related Work}
\label{sec:related}

The quality of experience (QoE) of video streaming over wireless networks has attracted a lot of attention recently.
Different from traditional quality of service (QoS) metrics, such as average throughput and delay as well as delay jitter,
QoE provides more direct measure to the human perception. Generally speaking, streaming QoE are interpreted in two ways,
objective QoE and subjective QoE. The former defines a set of metrics that can be exactly and objectively quantified in which
the start-up delay and starvation behaviors are the most prominent. Whereas the latter
is the direct grading by a user. The subjective QoE evaluation is usually intrusive, which is not scalable to a large population.
To solve this difficulty, authors in \cite{dobrian:sigcomm2011} adopted video abandonment rate as a 
measure to quantify the subjective QoE.
They performed a large scale measurement of view records and found that streaming QoE was mainly hurt by the percentage
of time wasted in the playback interruption. Authors in \cite{Balachandran2013Developing} presented a data-driven approach to model the relationship
between interdependent QoE metrics and user engagement, and proposed a predictive model for Internet video QoE.
Authors in \cite{Krishnan2013Video} studied the impact of video stream quality on viewer
behavior by using extensive traces from Akamai's streaming network.
These studies rely on client-side implementation to measure video quality metrics.
In \cite{Erman2011Over}, authors examined the video traffic generated by three million users in a large 3G cellular network.
Their focus was to infer the average video bitrate using deep packet inspection. Authors in \cite{dobrian:sigcomm2011}
present a large-scale study characterizing the impact of cellular network performance on mobile video
user engagement from the perspective of a network operator.

Although the user engagement is proven to be an efficient subjective measure of the objective QoE metrics,
the intrinsic relationship between objective QoE metrics and QoS metrics remains mysterious.
There exist a string of studies aiming to model the start-up delay and buffer starvation behaviors with different
throughput processes. Stockhammer et al. \cite{Stockhammer:streaming2002} specified the minimum start-up delay and the minimum buffer size
for a given video stream and a deterministic variable bit rate (VBR) wireless channel. Liang et al. \cite{Liang:multimedia2008}
considered a particular wireless channel model where the channel oscillates between good and bad states
following the extended Gilbert model. ParandehGheibi et al. \cite{A:interruptions2011} studied a particular P2P video streaming based
on random linear network coding, which simplified the packet requests at the network layer and allowed to
model the receiver buffer as an M/D/1 queue.
Authors in \cite{Luan:multimedia2010} studied a G/G/1 queue whose arrival and service rates are characterized by the first and second moments.
By using diffusion approximation, they obtained the closed-form starvation probability with asymptotically large file size.
Xu et al. \cite{XU:infocom2012} modeled the starvation probability for a file of finite size in M/M/1 and M/D/1 queues. Two approaches, the
Ballot theorem approach and the recursive approach, are employed to derive the probability generating function of
buffer starvations. They further extended this model to analyze the QoE metrics of a persistent video streaming in cellular networks.
In addition, network capacity is shared by the competing flows. Available capacity of each flow depends
on the amount of active flows, which depends on the arrival and departure of flows, so-called flow dynamics. Authors in \cite{xu:infocom2013}
proposed a set of mathematical models to characterize the distribution of start-up delay and buffer starvations when multiple
video streams dynamically share the same base station.

This paper falls in the scope of modeling QoE metrics, but differs from the literature in three aspects.
Firstly, we explore the fundamental role of network operators that can play to enhance streaming QoE, other than finding a
prefetching strategy to balance the tradeoff between the start-up delay and starvation probability.
Secondly, our work goes beyond the general measurement studies in that we present a novel and accurate
viewing time model, and apply it to compute QoE metrics. Thirdly, our purpose is to make our QoE models be computed by
network operators in practice. A simple Bayesian inference approach is proposed to enable content providers to release
``minimum'' information to network operators, while this information exchange has not been investigated so far.

\section{Conclusion}
\label{sec:conclusion}

In this work, We perform a large-scale measurement on the user behavior of Internet video streaming service.
For the first time we
find that the viewing time follows a hyper-exponential distribution. Assisted by this observation,
we propose a set of closed-form models for QoE metrics in a bandwidth shared wireless network, with the consideration
of DPS scheduling adopted by the base station.
The basic models are further generalized to incorporate a variety of practical issues including the progressive downloading
and the finite video duration. Trace-driven simulations validate the accuracy of our models.
Our study provides an important insight on the improvement of overall QoE, that is,
by favoring a 2-times higher scheduling weight to short views, they may experience a much smaller
risk of playback starvation, while the starvation probability of long views only increases slightly.
To the goal of differentiating short and long views, we design a novel approach to identify the class of an incoming flow
with \emph{minimum} information required by the network operators.

\bibliographystyle{IEEEtran}

\newpage
.
\newpage

\appendices
\section{The derivation of $\mathbf V(0;q_a)$}
Define the probability that startup phase starts at state \((i,j)\) and ends at state \((i',j')\)  as
$$
V_{i,j} ^{i',j'} (q;q_a):= \mathbb{P}\{I(T_a)=(i',j')|I(0)=(i,j), Q(0)=q\}
$$
where \(T_a\) is the ending time of startup.  Since tagged flow won't depart in startup phase, there is no absorbing state A. Besides, video is not played during startup phase,  so in the time internal \([0,h]\), the buffer length changes on the basis of the rule
\begin{equation}
Q(t+h)=Q(t)+b_{i,j}h
\end{equation}
Using the same approach of deriving $\mathbf W(q_a)$, there exists for all
\begin{align}
&V_{i,j} ^{i',j'}(q;q_a)=(1-(\lambda+\mu_{i,j}+\nu_{i,j})h)V_{i,j} ^{i',j'} (q+b_{i,j}h;q_a)\nonumber \\
&+\lambda_1hV_{i+1,j} ^{i',j'} (q+b_{i,j}h;q_a)+\lambda_2hV_{i,j+1} ^{i',j'} (q+b_{i,j}h;q_a) \nonumber\\
&+\mu_{i,j}hV_{i-1,j} ^{i'j'} (q+b_{i,j}h;q_a)+\nu_{i,j}hV_{i,j-1} ^{i',j'} (q+b_{i,j}h;q_a) \nonumber \\
&+o(h)
\label{eq.v1}
\end{align}

When \(h\to0\), Eq.\eqref{eq.v1} yields the following set of ordinary differential equations (ODEs)
\begin{align}
c_{i,j}\dot V_{i,j} ^{i',j'}(q;q_a)&=(\lambda+\mu_{i,j}+\nu_{i,j})V_{i,j} ^{i',j'} (q;q_a)\nonumber \\
&+\lambda_1V_{i+1,j} ^{i',j'} (q;q_a)+\lambda_2V_{i,j+1} ^{i',j'} (q;q_a) \nonumber\\
&+\mu_{i,j}V_{i-1,j} ^{i'j'} (q;q_a)+\nu_{i,j}V_{i,j-1} ^{i',j'} (q;q_a)
\label{eq.v2}
\end{align}
with the boundary condition
\begin{equation}
V_{i,j} ^{i',j'}(q_a;q_a) = \left\{
\begin{array}{rcl}
1      &      & \text{if }  (i,j)=(i',j');\\
0 &      & \text{otherwise.}
\end{array} \right.
\label{eq.v3}
\end{equation}

Similarly, let $\mathbf{V}(q;q_a)$ be a matrix of transition probability from different initial state to different terminative states, then Eq.\eqref{eq.v2} can be rewritten into a matrix form as
\begin{equation}
 \mathbf{\dot V}(q;q_a)= \mathbf M_V \mathbf{V}(q;q_a)
 \label{eq.vm}
\end{equation}
The matrix $\mathbf M_V$ is
\begin{equation}
\left(\begin{array}{ccccc}
    \mathbf S'_0 & \mathbf L'_0 & & & \\
    \mathbf U'_1 & \mathbf S'_1 & \mathbf L'_1 & & \\
        &\ddots&\ddots&\ddots& \\
        &       &\mathbf U'_{N_1(0)-1}&\mathbf S'_{N_1(0)-1}&\mathbf L'_{N_1(0)-1}\\
        &       &            &\mathbf U'_{N_1(0)}&\mathbf S'_{N_1(0)}
        \end{array}\right)
\end{equation}
where

$$
\mathbf L'_i= \begin{pmatrix}\begin{smallmatrix}
         \frac{-\lambda_1}{b_{i,0}} & & \\
           &\ddots&\frac{-\lambda_1}{b_{i,N_2(i)}}\\
           0&\cdots&0
           \end{smallmatrix}\end{pmatrix},
\mathbf U'_i=\begin{pmatrix}\begin{smallmatrix}
      \frac{-\mu_{i,0}}{b_{i,0}} & & &0 \\
      &\ddots& &\vdots \\
      &&\frac{-\mu_{i,N_2(i)}}{b_{i,0}}&0
     \end{smallmatrix}\end{pmatrix}
$$\\
and
$$
\mathbf S'_i=\begin{pmatrix}\begin{smallmatrix}
       S'_i(0) & S_i^{'+}(0) & & & \\
    S_i^{'-}(1) & S'_i(1) & S_i^{'+}(1) & & \\
        &\ddots&\ddots&\ddots& \\
        &       &            &S_i^{'-}(N_1(0))&S'_i(N_1(0))
      \end{smallmatrix}\end{pmatrix}
$$
with $S'_i(j)= \frac{\lambda+\mu_{i,j}+\nu_{i,j}}{b_{i,j}}$; $S_i^{'-}(j)=\frac{-\nu_{i,j}}{b_{i,j}}$ and $S_i^{'+}(j)=\frac{-\lambda_2}{b_{i,j}}$. The size of $\mathbf S'_i$ is $(K-i)\times (K-i)$, the size of $\mathbf L'_i$ is $(K-i+1)\times (K-i)$ and the size of $\mathbf U'_i$ is $(K-i)\times (K-i+1)$.

Because $\mathbf M_V$ is similar to a diagonal matrix, $\mathbf M_V$ can be eigendecomposed as $\mathbf M_V=\mathbf D_V\mathbf \Lambda_V \mathbf D_V^{-1}$, then the solution to Eq.\eqref{eq.vm} is
\begin{equation}
\mathbf V(q;q_a)= \mathbf D_V\exp(\mathbf \Lambda_Vq)\mathbf D_V^{-1}\cdot \mathbf V(0;q_a)
\label{eq.v4}
\end{equation}
Submitting Eq.\eqref{eq.v3} to Eq.\eqref{eq.v4}, we can have the solution as follows
\begin{equation}
\mathbf V(0;q_a)= \mathbf D_V\exp(-\mathbf \Lambda_Vq)\mathbf D_V^{-1}\cdot \mathbf V(q_a;q_a)
\end{equation}

\end{document}